\documentclass[
superscriptaddress,aps,preprintnumbers,amsmath,amssymb,prd,nofootinbib,preprint]{revtex4-1}

\usepackage{graphicx}
\usepackage{epstopdf}
\usepackage{dcolumn}
\usepackage{bm}
\usepackage{hyperref}
\usepackage{color}
\usepackage{amsmath}
\usepackage{multirow}

\begin{document}


\def\a{\alpha}
\def\b{\beta}
\def\c{\varepsilon}
\def\d{\delta}
\def\e{\epsilon}
\def\f{\phi}
\def\g{\gamma}
\def\h{\theta}
\def\k{\kappa}
\def\l{\lambda}
\def\m{\mu}
\def\n{\nu}
\def\p{\psi}
\def\q{\partial}
\def\r{\rho}
\def\s{\sigma}
\def\t{\tau}
\def\u{\upsilon}
\def\v{\varphi}
\def\w{\omega}
\def\x{\xi}
\def\y{\eta}
\def\z{\zeta}
\def\D{\Delta}
\def\G{\Gamma}
\def\H{\Theta}
\def\L{\Lambda}
\def\F{\Phi}
\def\P{\Psi}
\def\S{\Sigma}

\def\o{\over}
\def\beq{\begin{align}}
\def\eeq{\end{align}}
\newcommand{\gsim}{ \mathop{}_{\textstyle \sim}^{\textstyle >} }
\newcommand{\lsim}{ \mathop{}_{\textstyle \sim}^{\textstyle <} }
\newcommand{\vev}[1]{ \left\langle {#1} \right\rangle }
\newcommand{\bra}[1]{ \langle {#1} | }
\newcommand{\ket}[1]{ | {#1} \rangle }
\newcommand{\EV}{ {\rm eV} }
\newcommand{\KEV}{ {\rm keV} }
\newcommand{\MEV}{ {\rm MeV} }
\newcommand{\GEV}{ {\rm GeV} }
\newcommand{\TEV}{ {\rm TeV} }
\newcommand{\1}{\mbox{1}\hspace{-0.25em}\mbox{l}}
\newcommand{\headline}[1]{\noindent{\bf #1}}
\def\diag{\mathop{\rm diag}\nolimits}
\def\Spin{\mathop{\rm Spin}}
\def\SO{\mathop{\rm SO}}
\def\O{\mathop{\rm O}}
\def\SU{\mathop{\rm SU}}
\def\U{\mathop{\rm U}}
\def\Sp{\mathop{\rm Sp}}
\def\SL{\mathop{\rm SL}}
\def\tr{\mathop{\rm tr}}
\def\mpl{M_{\rm Pl}}

\def\IJMP{Int.~J.~Mod.~Phys. }
\def\MPL{Mod.~Phys.~Lett. }
\def\NP{Nucl.~Phys. }
\def\PL{Phys.~Lett. }
\def\PR{Phys.~Rev. }
\def\PRL{Phys.~Rev.~Lett. }
\def\PTP{Prog.~Theor.~Phys. }
\def\ZP{Z.~Phys. }

\def\dd{\mathrm{d}}
\def\ff{\mathrm{f}}
\def\BH{{\rm BH}}
\def\inf{{\rm inf}}
\def\ev{{\rm evap}}
\def\eq{{\rm eq}}
\def\SM{{\rm sm}}
\def\Mpl{M_{\rm Pl}}
\def\GeV{{\rm GeV}}
\newcommand{\Red}[1]{\textcolor{red}{#1}}
\newcommand{\TL}[1]{\textcolor{blue}{\bf TL: #1}}


\title{
750 GeV Diphotons:\\
\vspace{1mm}
Implications for Supersymmetric Unification II
}

\author{Lawrence J.~Hall}
\affiliation{Department of Physics, 
 University of California, Berkeley, California 94720, USA}
\affiliation{Theoretical Physics Group, 
 Lawrence Berkeley National Laboratory, Berkeley, California 94720, USA}
\author{Keisuke Harigaya}
\affiliation{Department of Physics, 
 University of California, Berkeley, California 94720, USA}
\affiliation{Theoretical Physics Group, 
 Lawrence Berkeley National Laboratory, Berkeley, California 94720, USA}
\author{Yasunori Nomura}
\affiliation{Department of Physics, 
 University of California, Berkeley, California 94720, USA}
\affiliation{Theoretical Physics Group, 
 Lawrence Berkeley National Laboratory, Berkeley, California 94720, USA}

\begin{abstract}

Perturbative supersymmetric gauge coupling unification is possible in six theories where complete $SU(5)$ TeV-scale multiplets of vector matter account for the size of the reported $750~{\rm GeV}$ diphoton resonance, interpreted as a singlet multiplet $S = (s+ia)/\sqrt{2}$.  One of these has a full generation of vector matter and a unified gauge coupling $\alpha_G \sim 1$.  The diphoton signal rate is enhanced by loops of vector squarks and sleptons, especially when the trilinear $A$ couplings are large.  If the $SH_u H_d$ coupling is absent, both $s$ and $a$ can contribute to the resonance, which may then have a large apparent width if the mass splitting from $s$ and $a$ arises from loops of vector matter. The width depends sensitively on $A$ parameters and phases of the vector squark and slepton masses.  Vector quarks and/or squarks are expected to be in reach of the LHC.  If the $SH_uH_d$ coupling is present, $a$ leads to a narrow diphoton resonance, while a second resonance with decays $s \rightarrow hh, W^+ W^-,ZZ$ is likely to be discovered at future LHC runs. In some of the theories a non-standard origin or running of the soft parameters is required, for example involving conformal hidden sector interactions.

\end{abstract}


\maketitle

\section{Introduction}

Data from both ATLAS and CMS experiments show evidence for a diphoton resonance near $750~{\rm GeV}$ \cite{ATLAS,CMS,ATLAS-2,CMS-2}.  We have previously explored the consistency of this data with perturbative gauge coupling unification in supersymmetric theories by adding a singlet field $S$ and vector matter $(\Phi_i, \bar{\Phi}_i)$ to the minimal supersymmetric standard model (MSSM)~\cite{Hall:2015xds} via the superpotential interaction $\lambda_i S \Phi_i \bar{\Phi}_i$.  A sufficient diphoton signal results only if $\lambda_i$ take values close to the maximum allowed by perturbativity, and hence we take them to be determined by renormalization group flow, yielding a highly predictive theory. The diphoton resonance has been further explored in this minimal supersymmetric theory~\cite{Nilles:2016bjl} as well as in other supersymmetric theories involving a singlet with vector matter in complete unified multiplets~\cite{Tang:2015eko,Ko:2016lai,Karozas:2016hcp,Dutta:2016jqn,King:2016wep,Han:2016fli,Barbieri:2016cnt}.

In this paper we further explore the diphoton resonance in minimal supersymmetric theories.  In addition to $\lambda_i S \Phi_i \bar{\Phi}_i$ we allow for the interaction of $S=(s+ia)/\sqrt{2}$ with Higgs doublets via $\lambda_H S H_u H_d$, giving the next-to-minimal supersymmetric standard model (NMSSM) with vector matter.  This additional interaction makes significant changes to the phenomenology, mixing $s$ with the doublet Higgs boson $h$ so that there is a further resonance to be discovered at the LHC of $s$ decaying to pairs of Higgs bosons or electroweak gauge bosons: $s \rightarrow hh, W^+ W^-,ZZ$. In this case the diphoton resonance is produced by $a$ alone, and is narrow.

As in Ref.~\cite{Hall:2015xds} we consider the complete set of 6 possibilities for vector matter that fills $SU(5)$ multiplets and allows perturbative gauge coupling unification: ``$({\bf 5} + \overline{\bf 5})_{N_5}$'' theories contain $N_5 = 1,2,3$ or $4$ copies of vector 5-plets, the ``${\bf 10} + \overline{\bf 10}$'' theory contains a single vector 10-plet, and the ``${\bf 15} + \overline{\bf 15}$'' theory contains a full generation of vector quarks and leptons.  In fact without threshold corrections the $({\bf 5} + \overline{\bf 5})_4$ and ${\bf 15} + \overline{\bf 15}$ theories become non-perturbative just before the gauge couplings unify.  We include these theories and study the form of the threshold corrections required to allow precision perturbative gauge coupling unification.  Indeed we find the ${\bf 15} + \overline{\bf 15}$ theory to be particularly interesting: supersymmetric theories with 4 or less generations have gauge couplings $\alpha_a$ much less than unity at the unification scale, while those with 6 or more generations become non-perturbative far below the unification scale.  The case of 5 generations, here interpreted as three chiral generations and one vector generation, is unique, offering the possibility of $\alpha_a \sim 1$ at the unification scale.

We compute the contribution to the diphoton signal from loops containing scalar superpartners of $(\Phi_i, \bar{\Phi}_i)$; such contributions were ignored in Ref.~\cite{Hall:2015xds} but were studied for 4 of our 6 theories in Ref.~\cite{Nilles:2016bjl}.  For each theory, the rates are computed for two cases corresponding to whether the supersymmetric mass terms of the vector matter, $\mu_i$, satisfy unified boundary conditions.  The corrections from the scalar loops become substantial and are important for large $A_i$ terms.  This is particularly important for the case of unified mass relations for $\mu_i$, since in the absence of the scalar contributions the rates are frequently marginal or inadequate to explain the data.  For example, the ${\bf 15} + \overline{\bf 15}$ theory with unified mass relations is only viable with large $A_i$.  Although the scalar mass parameters introduce further parameters, the unification of $\mu_i$ reduces the parameter space.  

In general, contributions from multiplets $(\Phi_i, \bar{\Phi}_i)$ to the diphoton amplitude add with random phases, or random signs if CP is conserved, typically significantly   reducing the signal rate.  We introduce theories where the mass terms for the vector matter arise purely from a condensate of $S$, giving $\mu_i \sim \lambda_i \vev{S}$, which has the effect of aligning the amplitudes from each multiplet and maximizing the signal rate. In addition, the resulting values for $\mu_i$ correspond to the case of unified masses.  Thus while the theories become more predictive, large $A_i$ are needed in some theories for a sufficient signal rate.

We explore the possibility that the mass splitting between the two scalar degrees of freedom in $S$ arise from loops containing $(\Phi_i, \bar{\Phi}_i)$.  It was argued in Ref.~\cite{Hall:2015xds} that when $\lambda_H =0$ such splittings could lead to an apparent width of 10s of GeV for the diphoton resonance.  Here we extend the analysis to include $A_i$ terms as well as CP violation in the holomorphic scalar mass terms of $(\Phi_i, \bar{\Phi}_i)$.  

We order our analysis as follows.  In the next section we compute the diphoton rate, with separate subsections for the cases of $\lambda_H=0$ and $\lambda_H \neq 0$.  In the latter case, in addition to having Higgsino loop contributions, the diphoton rate arises from only one scalar mode of $S$, as the other mixes with the light Higgs boson.  In section~\ref{sec:width} we discuss the width of the resonance for $\lambda_H=0$.  In section~\ref{sec:other scalar} we switch to $\lambda_H \neq 0$ and study the diboson LHC signal that results from one component of $S$ mixing with the light Higgs boson.  The condition on threshold corrections for perturbative unification in $({\bf 5} + \overline{\bf 5})_4$ and ${\bf 15} + \overline{\bf 15}$ theories is studied in section~\ref{sec:running} and theories with $\mu_i \sim \lambda_i \vev{S}$ are introduced in section~\ref{sec:Svev}.

\section{750 GeV diphoton resonance}
\label{sec:rate}

In this section, we discuss an explanation of the diphoton excess observed at the LHC~\cite{ATLAS,CMS,ATLAS-2,CMS-2}.  We introduce a singlet chiral multiplet $S$ and pairs of $SU(5)$ charged chiral multiplets $\Phi_i$ and $\bar{\Phi}_i$ around the TeV scale, and take the most general superpotential couplings and mass terms
\begin{align}
W \supset  S \sum_i \lambda_i \Phi_i \overline{\Phi}_i  + \lambda_H S H_u H_d+ \sum_i\mu_i\Phi_i \overline{\Phi}_i  + \mu_H H_u H_d + \frac{\mu_S}{2}S^2  +\frac{\kappa}{6}S^3. 
\label{eq:superpotential}
\end{align}
The coupling $\kappa$ flows to small values at low energies and is unimportant for the analysis of this paper.  In section~\ref{sec:Svev} we briefly mention its possible role in stabilizing a vacuum expectation value (vev) for $S$.  We consider the complete set of possible theories with perturbative gauge coupling unification: the ``$({\bf 5} + \overline{\bf 5})_{N_5}$'' theory containing $N_5 = 1,2,3$ or $4$ copies of $(\bar{D},\bar{L}) + (D,L)$, the ``${\bf 10} + \overline{\bf 10}$'' theory containing $(Q,U,E) + (\bar{Q},\bar{U},\bar{E})$, and the ``${\bf 15} + \overline{\bf 15}$'' theory that contains a full generation of vector quarks and leptons.  In the $({\bf 5} + \overline{\bf 5})_4$ and ${\bf 15} + \overline{\bf 15}$ theories, the standard model gauge couplings near the unification scale $M_G$ are in the strong coupling regime if all super particles are below $1~{\rm TeV}$.  We discuss the running of the gauge couplings and the threshold corrections around the TeV scale for these theories in section~\ref{sec:running}.

The diphoton signal is explained by the production of the scalar component(s) of $S$ via gluon fusion and the subsequent decay into diphotons, which are induced by the loop correction of $\Phi_i$ and $\bar{\Phi}_i$.  For $\lambda_H \neq0$, $s$ mixes with doublet Higgs and efficiently decays into a pair of standard model Higgs or gauge bosons, and does not contribute to the diphoton signal.  Thus, we consider the cases with $\lambda_H=0$ and $\lambda_H \neq 0$ independently.  For $\lambda_H \neq 0$, the LHC signal of $s \rightarrow hh, W^+ W^-,ZZ$ is discussed in section~\ref{sec:other scalar}.

\subsection{Vanishing Higgs coupling: {\boldmath $\lambda_H=0$}}

Let us first discuss the size of $\lambda_i$ and $\mu_i $.  As we have shown in Ref.~\cite{Hall:2015xds}, as long as $\lambda_i$ are large enough at high energies, they flow into quasi-fixed points and their low energy values are insensitive to the high energy values.  In Table~\ref{tab:coupling1}, we show the prediction for $\lambda_i({\rm TeV})$ in each theory, which we assume in the following.%
\footnote{The predicted values in $({\bf 5} + \overline{\bf 5})_4$ and ${\bf 15} + \overline{\bf 15}$ are different from those in Ref.~\cite{Hall:2015xds}.  In these theories, as we will see in section~\ref{sec:running}, the gauge coupling unification requires moderate threshold correction around the TeV scale, which is not taken into account in Ref.~\cite{Hall:2015xds}.  This changes the gauge couplings above the TeV scale as well as the predictions of $\lambda_i$.  If we instead assume large threshold corrections at the unification scale, the predictions of Ref.~\cite{Hall:2015xds} hold.}
If $\mu_i$ unify at the unification scale, their relative values are fixed by the renormalization group running, which are also shown in Table~\ref{tab:coupling1}.

\begin{table}[b]
\begin{center}
\begin{tabular}{|l|l||c|c|c|c|c|}
\hline
    & & $D$ & $L$ & $Q$ & $U$ & $E$ \\ \hline\hline
  \multirow{2}{*}{$({\bf 5} + \overline{\bf 5})_1$} & $\lambda_i$ 
    & 0.96 & 0.63 & \multirow{2}{*}{---} & \multirow{2}{*}{---} 
    & \multirow{2}{*}{---} \\
    & $\mu_i/\mu_L$ & 1.5 & 1 & & & \\ \hline
  \multirow{2}{*}{$({\bf 5} + \overline{\bf 5})_2$} & $\lambda_i$ 
    & 0.77 & 0.46 & \multirow{2}{*}{---} & \multirow{2}{*}{---} 
    & \multirow{2}{*}{---} \\
    & $\mu_i/\mu_L$ & 1.7 & 1 & & & \\ \hline
  \multirow{2}{*}{$({\bf 5} + \overline{\bf 5})_3$} & $\lambda_i$ 
    & 0.70 & 0.36 & \multirow{2}{*}{---} & \multirow{2}{*}{---} 
    & \multirow{2}{*}{---} \\
    & $\mu_i/\mu_L$ & 1.9 & 1 & & & \\ \hline
  \multirow{2}{*}{$({\bf 5} + \overline{\bf 5})_4$} & $\lambda_i$ 
    & 0.67 & 0.22 & \multirow{2}{*}{---} & \multirow{2}{*}{---} 
    & \multirow{2}{*}{---} \\
    & $\mu_i/\mu_L$ & 3.0 & 1 & & & \\ \hline
  \multirow{2}{*}{${\bf 10} + \overline{\bf 10}$} & $\lambda_i$ 
    & \multirow{2}{*}{---} & \multirow{2}{*}{---} & 0.87 & 0.71 & 0.26 \\
    &  $\mu_i/\mu_E$ &  &  & 3.0 & 2.5 & 1 \\ \hline
  \multirow{2}{*}{${\bf 15} + \overline{\bf 15}$} & $\lambda_i$ 
    & 0.60 & 0.17  & 0.85 & 0.64 & 0.12 \\
    & $\mu_i/\mu_E$ & 5.0 & 1.4 & 7.1 & 5.3 & 1 \\ \hline
\end{tabular}
\caption{Predictions for $\lambda_i$(TeV) and physical mass ratios $\mu_i/ \mu_{L,E}$ at one loop level assuming $\lambda_H = 0$.  The mass ratios assume a common value for $\mu_i$ at $M_G \simeq 2 \times 10^{16}~{\rm GeV}$.}
\label{tab:coupling1}
\end{center}
\end{table}

After taking $\mu_i$ to be real by phase rotations of $\Phi_i$ and $\bar{\Phi}_i$, $\lambda_i$ are in general complex.  We assume that $\lambda_i$ have a common phase.  This is automatic for $({\bf 5} + \overline{\bf 5})$, $({\bf 10} + \overline{\bf 10})$ and $({\bf 15} + \overline{\bf 15})$ theories with unified $\lambda_i$ and $\mu_i$ at the unification scale.%
\footnote{The alignment  is also guaranteed if $\mu_i$ are solely given by a vev of $S$. See section~\ref{sec:Svev}.}
By a phase rotation of $S$, we take $\lambda_i$ to be real and positive.
In this basis, we decompose the scalar components of $S$ as
\begin{align}
S = \frac{1}{\sqrt{2}} \left( s + i a \right),
\end{align}
and refer to ``$s$" and ``$a$" as ``scalar" and ``pseudoscalar", respectively.
They may be degenerate so that both contribute to the diphoton excess at $750~{\rm GeV}$, which we assume unless otherwise stated.   The mass splitting between the two scalars is discussed in section~\ref{sec:width}.

\begin{figure}[t]
\begin{center}
  \includegraphics[scale=0.6]{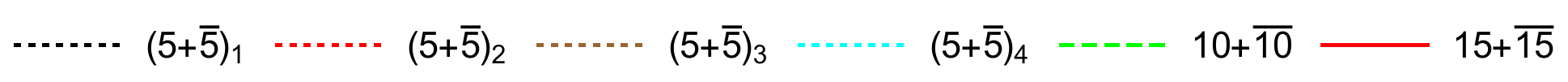}\\
  \vspace{0.5cm}
  \includegraphics[scale=0.6]{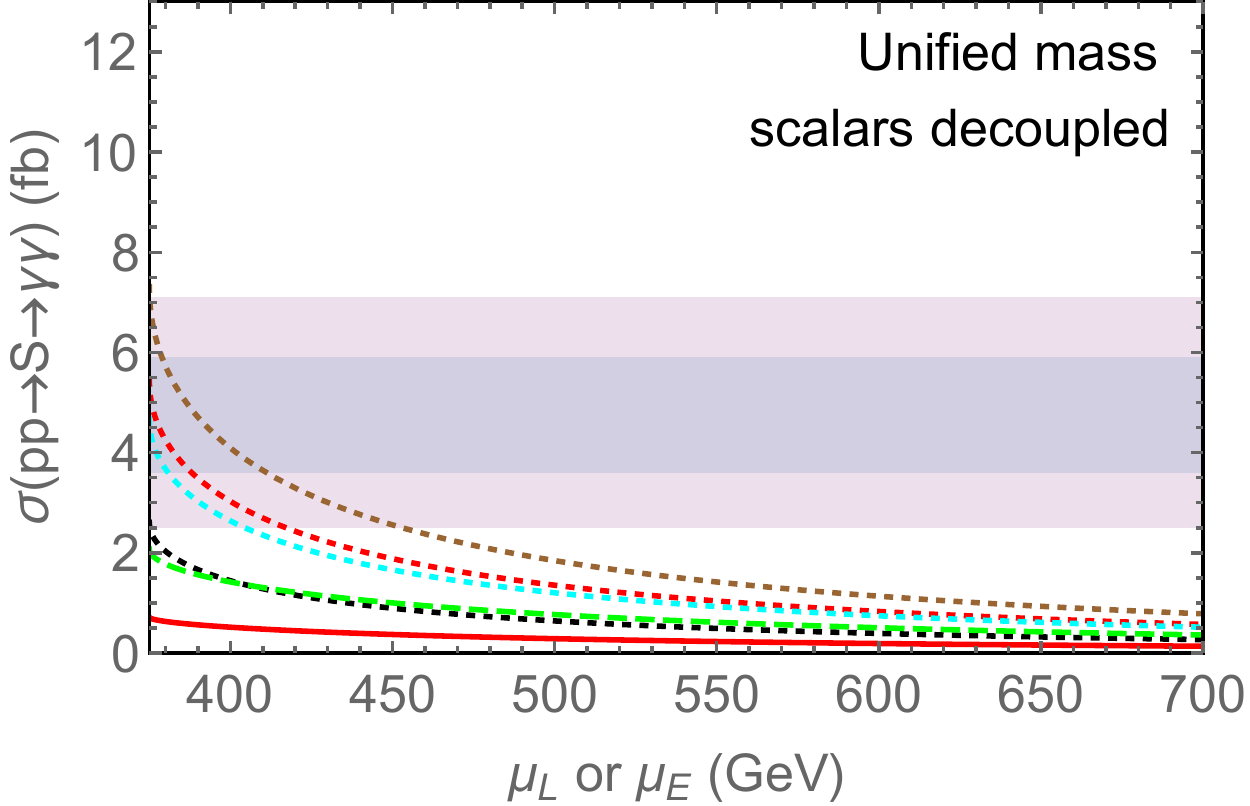}
    \includegraphics[scale=0.6]{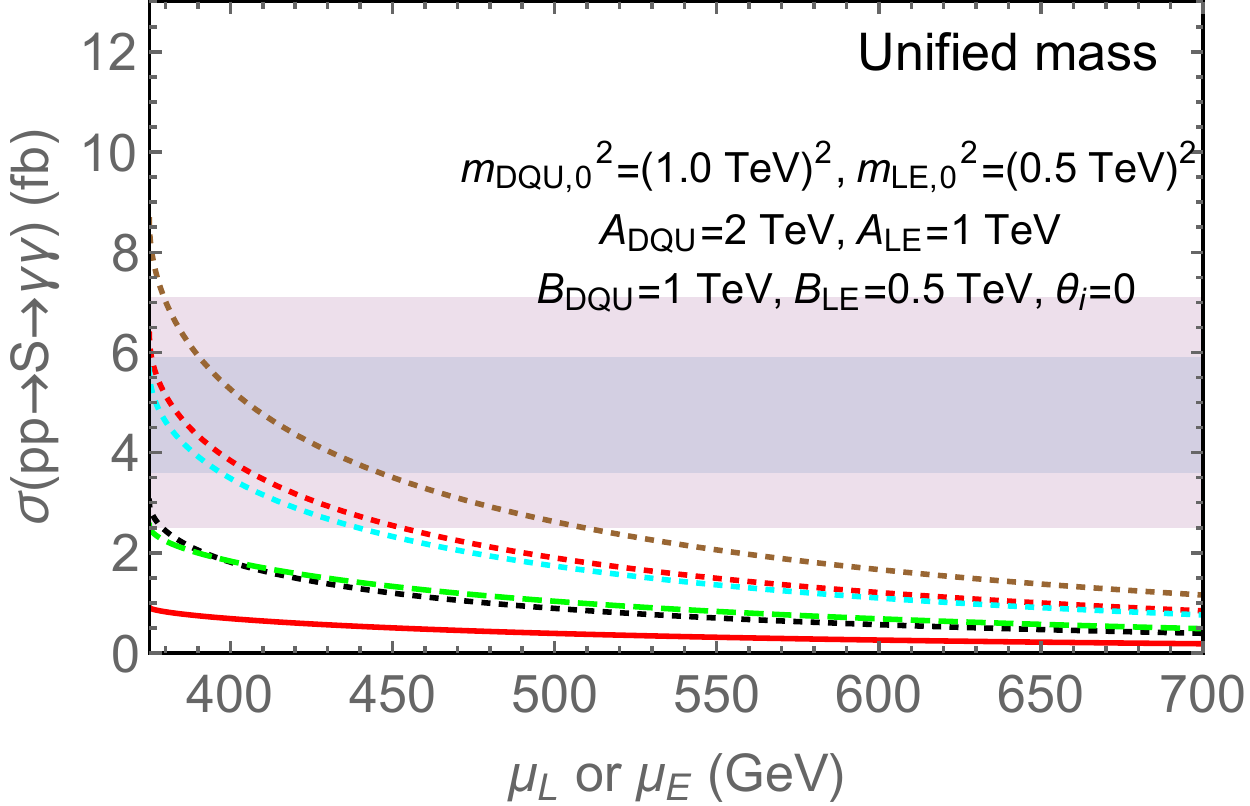}
    \includegraphics[scale=0.6]{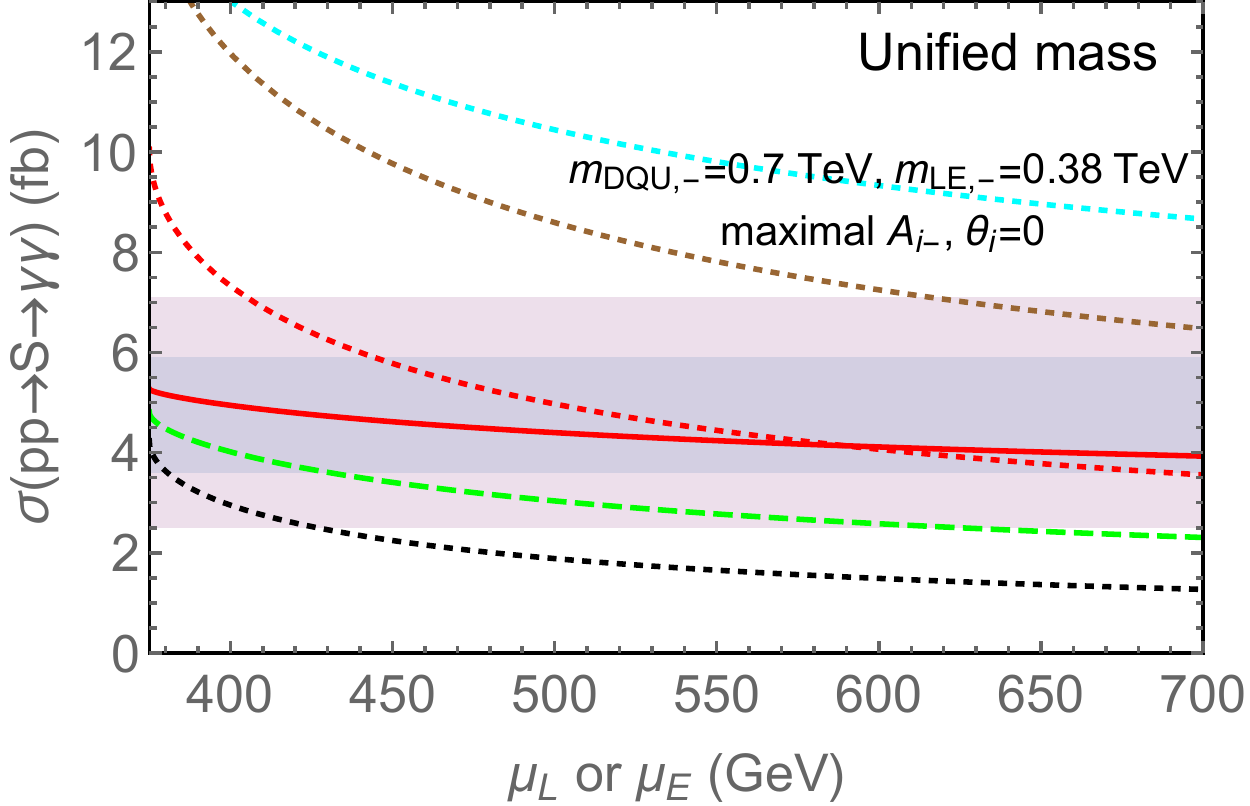}
        \includegraphics[scale=0.6]{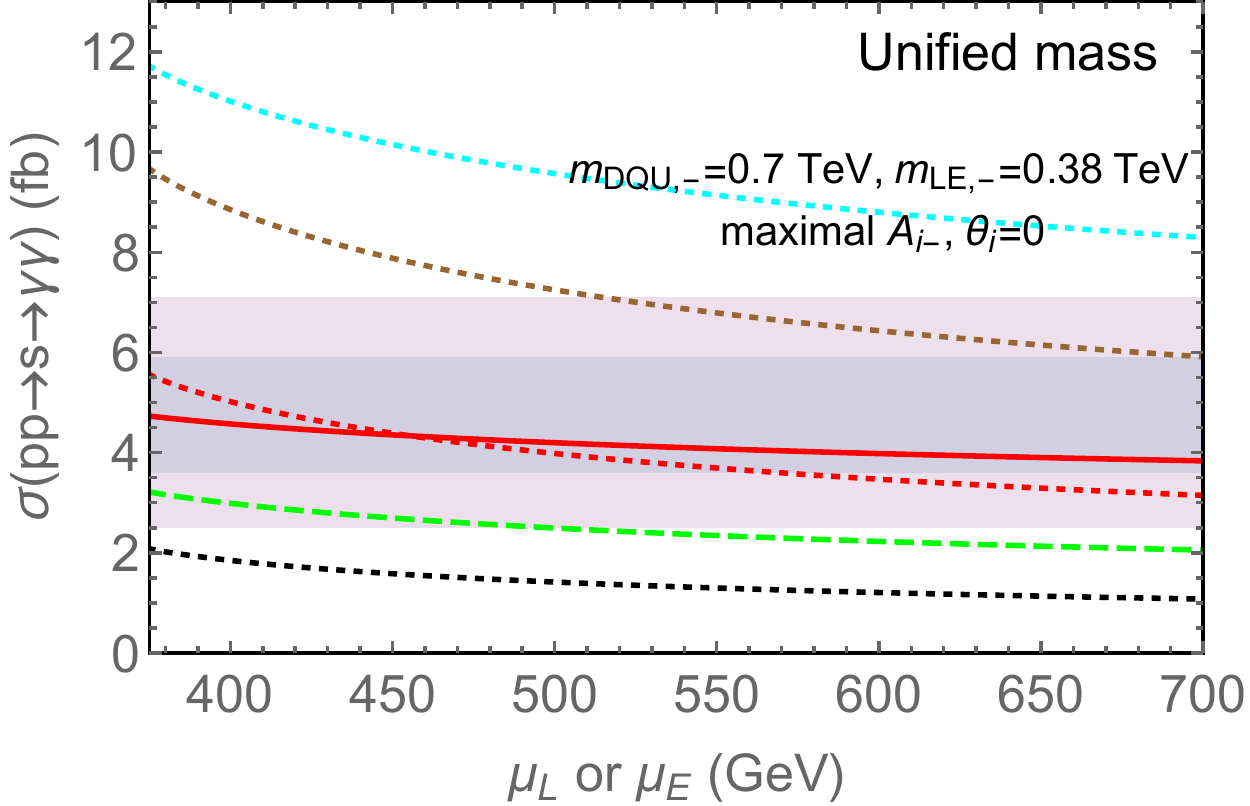}
\end{center}
\caption{Theories with unified mass relations and $\lambda_H=0$: Prediction for $\sigma B_{\gamma \gamma}$ at $\sqrt{s} = 13~{\rm TeV}$ as a function of the lightest vector lepton mass, with scalar partners decoupled (upper left), soft masses indicated in the figure (upper right), and the maximal possible $A_{i-}$ terms (lower left).  In the lower right panel, the contribution only from the scalar $s$ is depicted with the maximal possible $A_{i-}$ term.}
\label{fig:unify_twores}
\end{figure}

\begin{figure}[t]
\begin{center}
  \includegraphics[scale=0.6]{legend.pdf}\\
  \vspace{0.5cm}
  \includegraphics[scale=0.6]{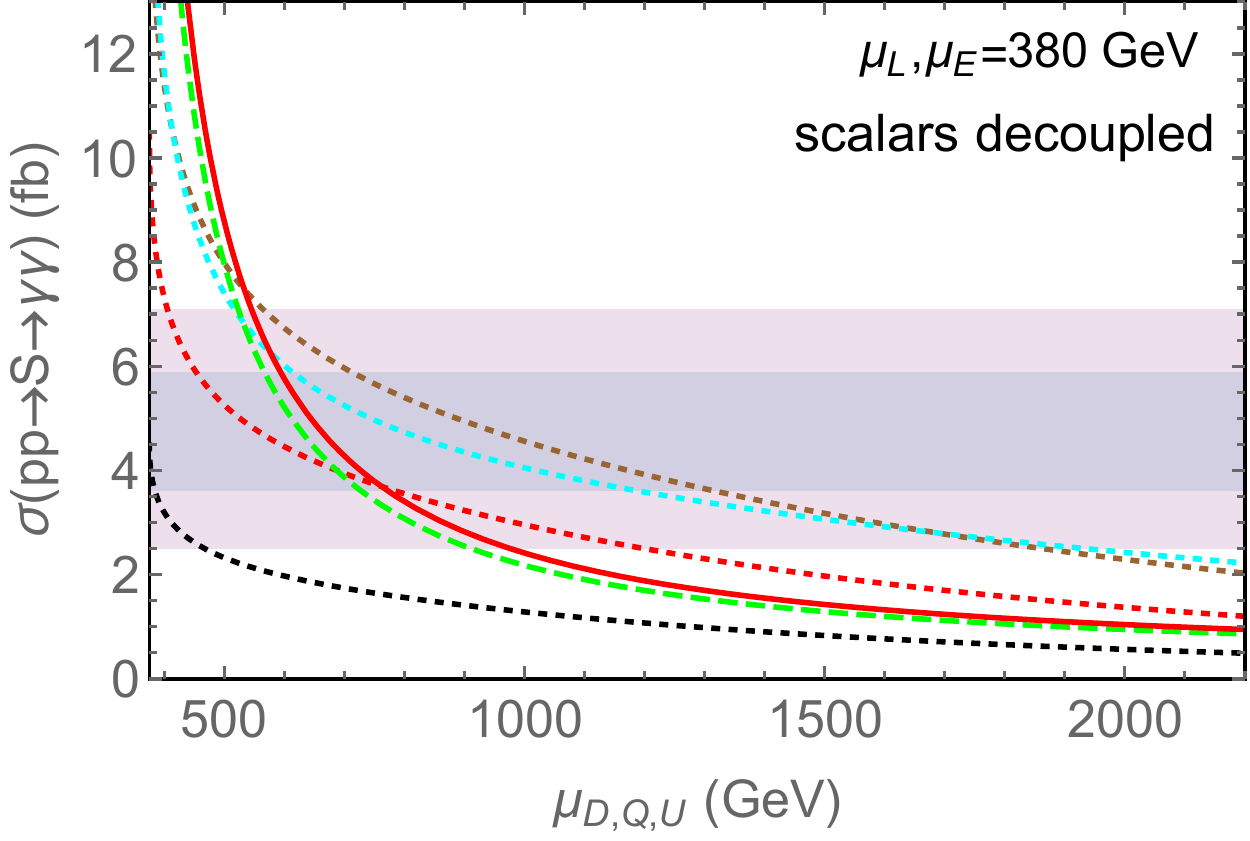}
    \includegraphics[scale=0.6]{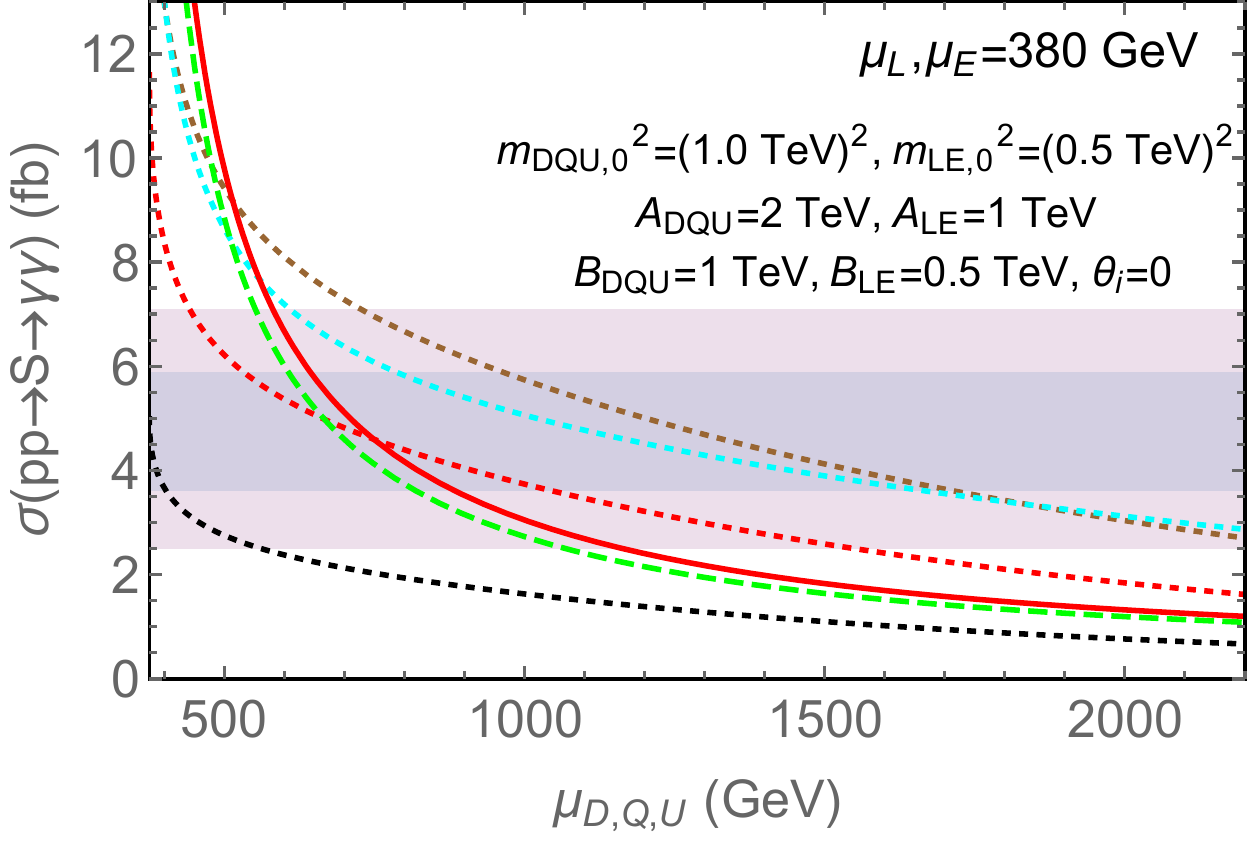}
    \includegraphics[scale=0.6]{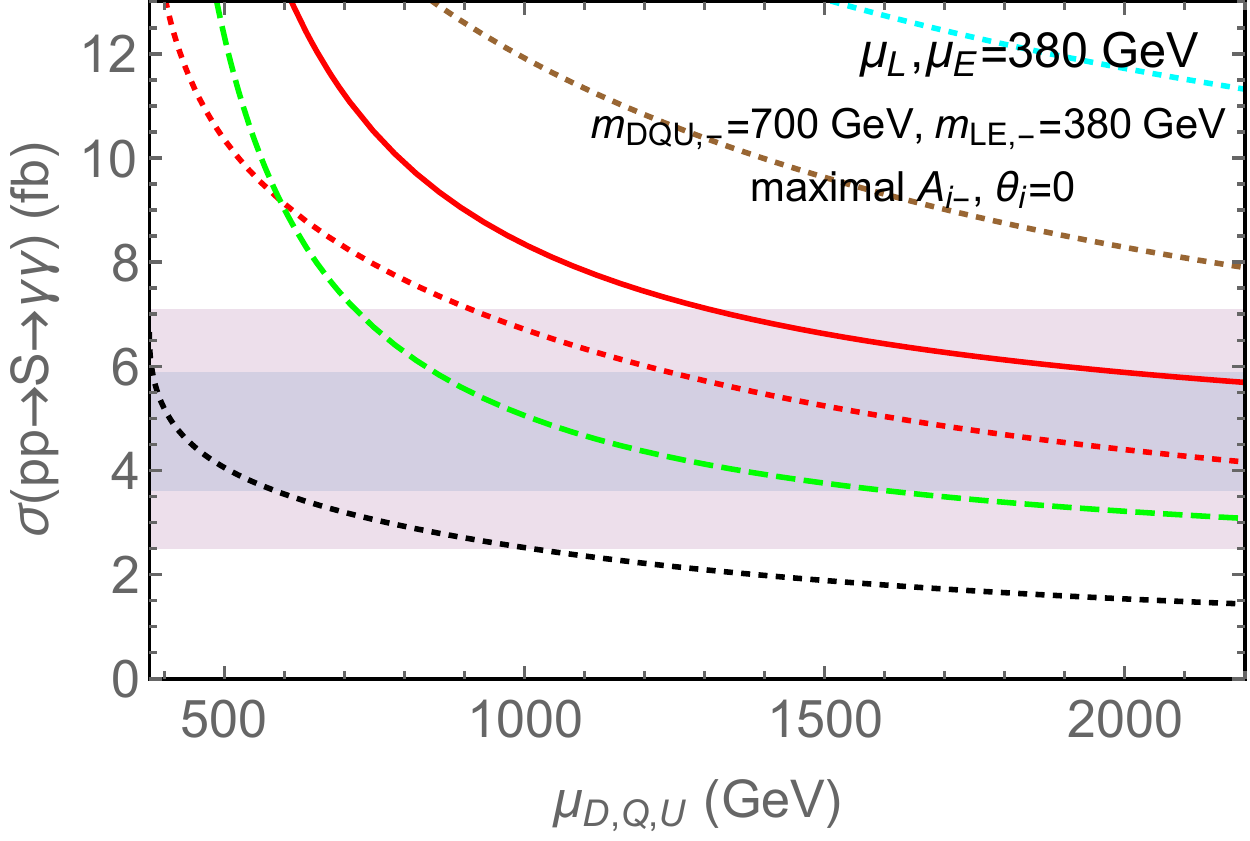}
        \includegraphics[scale=0.6]{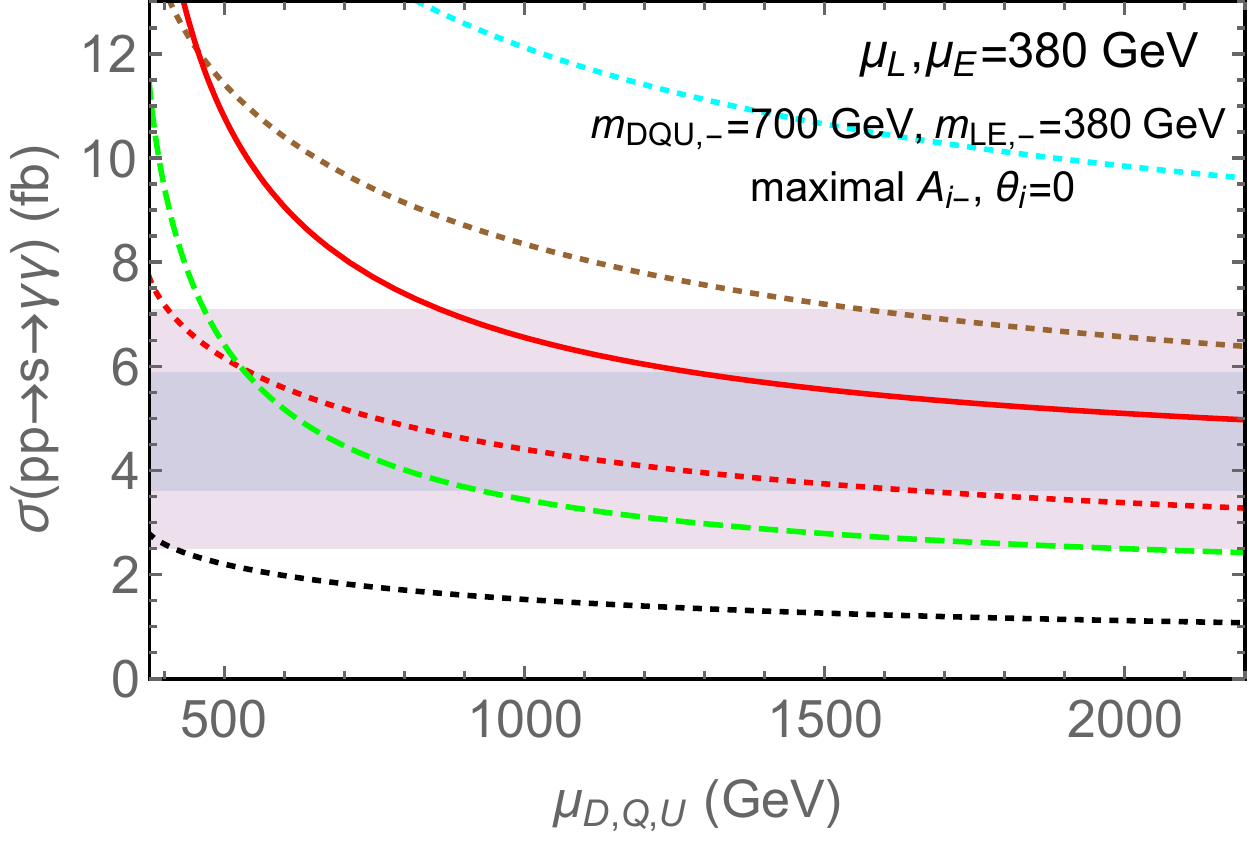}
\end{center}
\caption{Theories without unified mass relations and $\lambda_H=0$: Prediction for $\sigma B_{\gamma \gamma}$ at $\sqrt{s} = 13~{\rm TeV}$ as a function of the degenerate vector quark mass for vector lepton masses at $400~{\rm GeV}$, with scalar partners decoupled (upper left), soft masses indicated in the figure (upper right), and the maximal possible $A_{i-}$ terms (lower left).  In the lower right panel, the contribution from only the scalar $s$ is depicted with the maximal possible $A_{i-}$ term.}
\label{fig:optimal_twores}
\end{figure}

The upper left panel of Figure~\ref{fig:unify_twores}, shows the prediction for $\sigma_S {\rm Br}_{\gamma\gamma}$ at the LHC with $\sqrt{s}=13~{\rm TeV}$ as a function of $\mu_L$ for $({\bf 5} + \overline{\bf 5})_i$ and $\mu_E$ for $({\bf 10} + \overline{\bf 10})$ and $({\bf 15} + \overline{\bf 15})$, assuming that $\mu_i$ unify at the unification scale.  We also assume that  the scalar components of $\Phi_i$ and $\bar{\Phi}_i$ are heavy enough that their loop corrections do not contribute to the signal.  For this case, only the $({\bf 5} + \overline{\bf 5})_{2,3,4}$ theories can explain the observed diphoton excess and require light vector matter.  Vector quark masses are predicted in the range $700-1200~{\rm GeV}$, which can be observed at the LHC.  In Figures~\ref{fig:unify_twores}~--~\ref{fig:optimal_oneres} we show shaded 1$\sigma$ and 2$\sigma$ regions for a signal rate of $\sigma {\rm Br}_{\gamma\gamma} = (4.7+1.2 - 1.1)~{\rm fb}$ from combined fits to the experimental data~\cite{Buttazzo:2016kid}.

Once we relax the assumption of the unification of $\mu_i$ at the unification scale, the possibilities for explaining the $750~{\rm GeV}$ excess are greatly expanded.  This occurs in theories in which boundary conditions in extra dimensions break the unified symmetry~\cite{Hall:2001pg}. It can also occur in four dimensional theories if these masses pick up unified symmetry breaking effects at an $O(1)$ level.  In the upper left panel of Figure~\ref{fig:optimal_twores}, we show the prediction for $\sigma_S {\rm Br}_{\gamma\gamma}$ at the LHC with $\sqrt{s}=13~{\rm TeV}$ as a function of degenerate vector quark masses with vector lepton masses fixed at $\mu_{L,E}=380~{\rm GeV}$.  We again assume that the scalar components of $\Phi_i$ and $\bar{\Phi}_i$ are sufficiently heavy not to contribute.  Now all theories can explain the observed diphoton excess.  For $({\bf 5} + \overline{\bf 5})_{1}$, the masses of the vector quarks are as low as $500~{\rm GeV}$.  This satisfies the lower bound on the vector quark mass, if it dominantly decays into first or second generation quarks~\cite{Aad:2015tba}.  For $({\bf 5} + \overline{\bf 5})_{3,4}$, vector quark masses can be as large as $2~{\rm TeV}$.


Next, let us take into account the effect of the scalar components of $\Phi_i$ and $\bar{\Phi}_i$, which is also investigated in Ref.~\cite{Nilles:2016bjl}.  The trilinear couplings between the scalar components of $S$, $\Phi_i$ and $\bar{\Phi}_i$ are given by
\begin{align}
-{\cal L}_{\rm tri} =   \lambda_i \mu_i (S + S^*) \left( |\Phi_i|^2 + |\bar{\Phi}_i|^2  \right) + \left(  \lambda_i\left( \mu_S^* S^* + A_i S \right) \Phi_i \bar{\Phi}_i + {\rm h.c.}   \right),
\end{align}
where $A_i$ are soft trilinear couplings.  We take $A_i$ to be real by phase rotations of the scalar components of $\Phi_i$ and $\bar{\Phi}_i$, and we neglect the trilinear couplings proportional to $\mu_S$.  The mass terms of $\Phi_i$ and $\bar{\Phi}_i$ are given by
\begin{align}
V_{\rm mass} = m_{\Phi_i}^2 |\Phi_i|^2 + m_{\bar{\Phi}_i}^2 |\bar{\Phi}_i|^2  + \left( B_i \mu_i \Phi_i \bar{\Phi}_i + {\rm h.c.} \right).
\end{align}
Assuming $m_{\Phi_i}^2 = m_{\bar{\Phi}_i}^2 \equiv m_{i,0}^2 $, the mass eigenbasis, $(\Phi_{i+},\Phi_{i-})$, is given by
\begin{align}
\begin{pmatrix}
\Phi_i \\ \bar{\Phi}_i^*
\end{pmatrix} 
=
\begin{pmatrix}
\frac{1}{\sqrt{2}} & e^{-i \theta_{i} } \frac{1}{\sqrt{2}}\\
-e^{i \theta_{i} } \frac{1}{\sqrt{2}} & \frac{1}{\sqrt{2}}
\end{pmatrix}
\begin{pmatrix}
\Phi_{i+} \\ \Phi_{i-}
\end{pmatrix},
\end{align}
with masses
\begin{align}
m_{i\pm}^2 =  \mu_i^2 +  m_i^2 \pm  |B_i| \mu_i.
\end{align}
Here, $\theta_{i}$ is the phase of $B_i$, $B_i = e^{i\theta_{i}} |B_i|$.  The trilinear couplings in the mass eigenbasis are given by
\begin{align}
-{\cal L}_{\rm tri} &=
\frac{\lambda_i}{\sqrt{2}}\left(  A_{si+} s |\Phi_+|^2 +   A_{si-} s |\Phi_-|^2
+ A_{ai+} a |\Phi_+|^2 +   A_{ai-} a |\Phi_-|^2 \right), \\
& A_{si\pm} \equiv  \mp  A_i{\rm cos}\theta_i + 2 \mu_i ,~~
A_{ai \pm} \equiv  \mp  A_i {\rm sin}\theta_i,
\end{align}
where we neglect couplings proportional to $\Phi_+ \Phi_-^*$, which are irrelevant for the diphoton signal.

In the upper right panels of Figures~\ref{fig:unify_twores} and \ref{fig:optimal_twores}, we show the diphoton signal rate including the scalar loop contributions.  We take reference values of the soft masses shown in the figures, with moderate values of $A_i = (1,2)~{\rm TeV}$ for vector (leptons, quarks).  The bounds on the vector quark/lepton masses are relaxed typically by $100~{\rm GeV}$.  Larger $A_i$ can further relax the bound~\cite{Nilles:2016bjl}.  In the lower left panels of Figures~\ref{fig:unify_twores} and \ref{fig:optimal_twores}, we take the maximal $A_{i-}$ allowed by stability of the vacuum, $m_{DQU-}=700~{\rm GeV}$, $m_{LE-}=380~{\rm GeV}$, and decoupled $\Phi_+$.  For the size and derivation of the maximal $A_i$, see appendix~\ref{sec:A term} and Ref.~\cite{Nilles:2016bjl}.  All theories can explain the diphoton excess.  Note, however, that large $A_i$ typically generate a large mass splitting between $s$ and $a$ by quantum corrections (see section~\ref{sec:width}).  Both $s$ and $a$ can contribute to the diphoton signal at $750~{\rm GeV}$ because the phases $\theta_i$ allow cancellations in the mass splitting, although tuning is required for a narrow width of the $750~{\rm GeV}$ resonance.  Alternatively, in the lower right panels of Figures~\ref{fig:unify_twores} and \ref{fig:optimal_twores}, we assume that the masses of the scalar and the pseudoscalar are sufficiently split that only the scalar $s$ contributes to the $750~{\rm GeV}$ excess.  Even in this case, due to large $A_i$, all theories except $({\bf 5} + \overline{\bf 5})$ can explain the diphoton excess.

\subsection{Non-vanishing Higgs coupling: {\boldmath $\lambda_H \neq 0$}}

Let us now turn on the coupling between $S$ and the Higgs multiplet, $\lambda_H$.  The existence of $\lambda_H$ slightly changes the renormalization running of couplings.  In Table~\ref{tab:coupling2}, we show the prediction for $\lambda_i({\rm TeV})$ in each theory.  Here we assume that $\lambda_H$ is also large at a high energy scale.  The low energy couplings are slightly smaller than those in the theory with $\lambda_H=0$.

\begin{table}[b]
\begin{center}
\begin{tabular}{| l  | l ||c|c|c|c|c|c|}
\hline
      & & $D$ & $L$ & $Q$ & $U$ & $E$ & $H$ \\ \hline\hline
  \multirow{2}{*}{$({\bf 5} + \overline{\bf 5})_1$} &  $\lambda_i$ 
    & 0.90 & 0.57 & \multirow{2}{*}{---} & \multirow{2}{*}{---} 
    & \multirow{2}{*}{---} & 0.45 \\
    & $\mu_i/\mu_L$ & 1.6 & 1 & & & &  \\ \hline
  \multirow{2}{*}{$({\bf 5} + \overline{\bf 5})_2$} & $\lambda_i$ 
    & 0.74 & 0.43 & \multirow{2}{*}{---} & \multirow{2}{*}{---} 
    & \multirow{2}{*}{---} & 0.33 \\
    &  $\mu_i/\mu_L$ & 1.7 & 1 & & & & \\ \hline
  \multirow{2}{*}{$({\bf 5} + \overline{\bf 5})_3$} &  $\lambda_i$ 
    & 0.67 & 0.33 & \multirow{2}{*}{---} & \multirow{2}{*}{---} 
    & \multirow{2}{*}{---} & 0.24 \\
    & $\mu_i/\mu_L$ & 2.0 & 1 & & & & \\ \hline
  \multirow{2}{*}{$({\bf 5} + \overline{\bf 5})_4$} &  $\lambda_i$ 
    & 0.67 & 0.21 & \multirow{2}{*}{---} & \multirow{2}{*}{---} 
    & \multirow{2}{*}{---} & 0.17 \\
    &  $\mu_i/\mu_L$ & 3.2 & 1 & & & & \\ \hline
  \multirow{2}{*}{${\bf 10} + \overline{\bf 10}$} &   $\lambda_i$ 
    & \multirow{2}{*}{---} & \multirow{2}{*}{---} & 0.86 & 0.69 & 0.26 & 0.24 \\
    &  $\mu_i/\mu_E$ &  &  & 3.3 & 2.7 & 1 & \\ \hline
  \multirow{2}{*}{${\bf 15} + \overline{\bf 15}$} &   $\lambda_i$ 
    & 0.59 & 0.16  & 0.84 & 0.64 & 0.12 & 0.14 \\
    &  $\mu_i/\mu_E$ & 4.9 & 1.3 & 7.0 & 5.3 & 1 &  \\ \hline
\end{tabular}
\caption{Predictions for $\lambda_i$(TeV) and physical mass ratios 
 $\mu_i/ \mu_{L,E}$ at one loop level with $\lambda_H \neq 0$.
 The mass ratios assume a common 
 value for $\mu_i$ at $M_G$.}
\label{tab:coupling2}
\end{center}
\end{table}

The mixing between the Higgs multiplet and $S$ is as follows.  Assuming the decoupling limit, large ${\rm tan}\beta$, and the CP conservation in the couplings between $S$ and the Higgs multiplet, the mass eigenstate is approximately given by the heavy Higgs states $(H^0,A^0, H^\pm)$ composed of $H_d$, the singlet pseudoscalar $a$, and the mixture of the standard model like Higgs $h$ and the singlet scalar $s$.  (With $\theta_i \neq 0,\pi$, quantum corrections inevitably induce mixing between $s$ and $a$; see section~\ref{sec:width}. The mixing is suppressed for sufficiently large $m_s^2$.  The pseudo-scalar $a$ mixes with the heavy CP-odd Higgs $A^0$ through the $A$ term coupling between $S$ and the Higgs multiplet.  This leads to the decay of $a$ into a pair of bottom quarks. The decay mode does not affect the diphoton signal rate for a sufficiently large heavy Higgs mass, a sufficiently small $A$ term, and/or not very large ${\rm tan}\beta$.)

\begin{figure}[t]
\begin{center}
  \includegraphics[scale=0.6]{legend.pdf}\\
  \vspace{0.5cm}
  \includegraphics[scale=0.6]{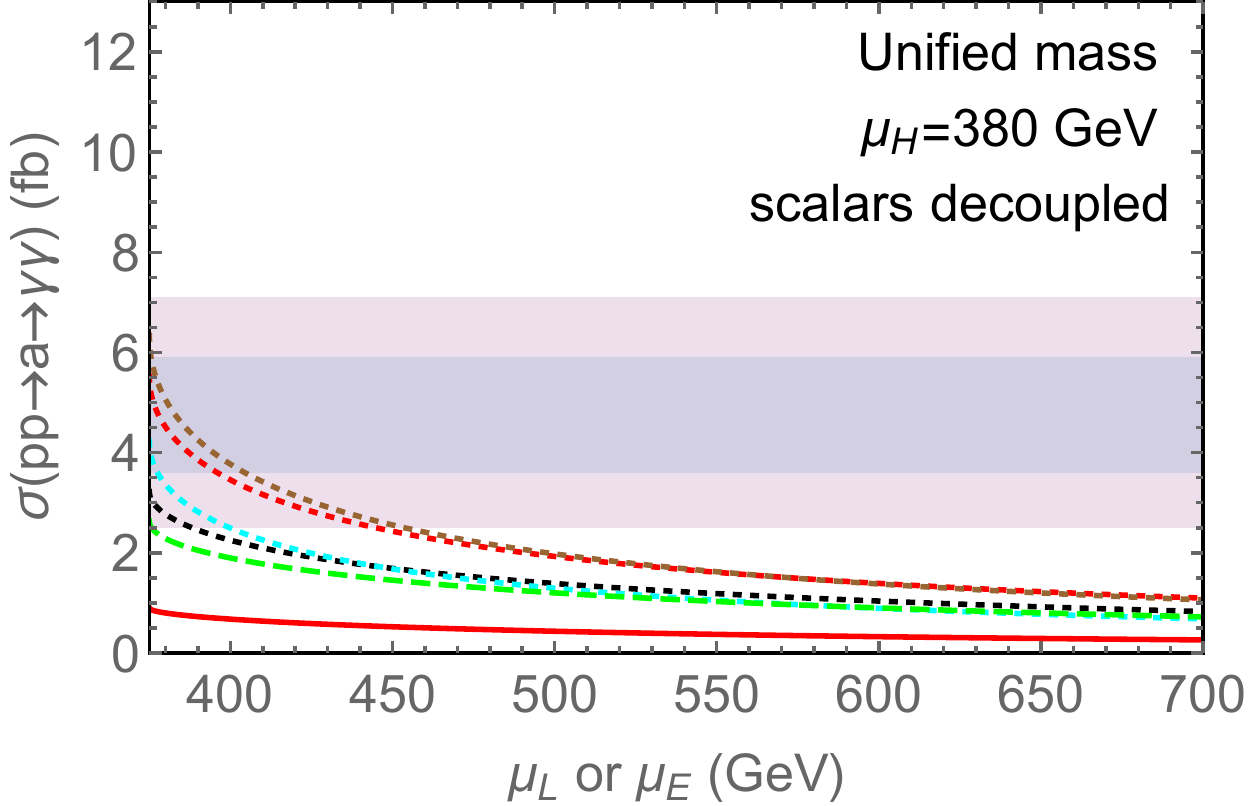}
  \includegraphics[scale=0.6]{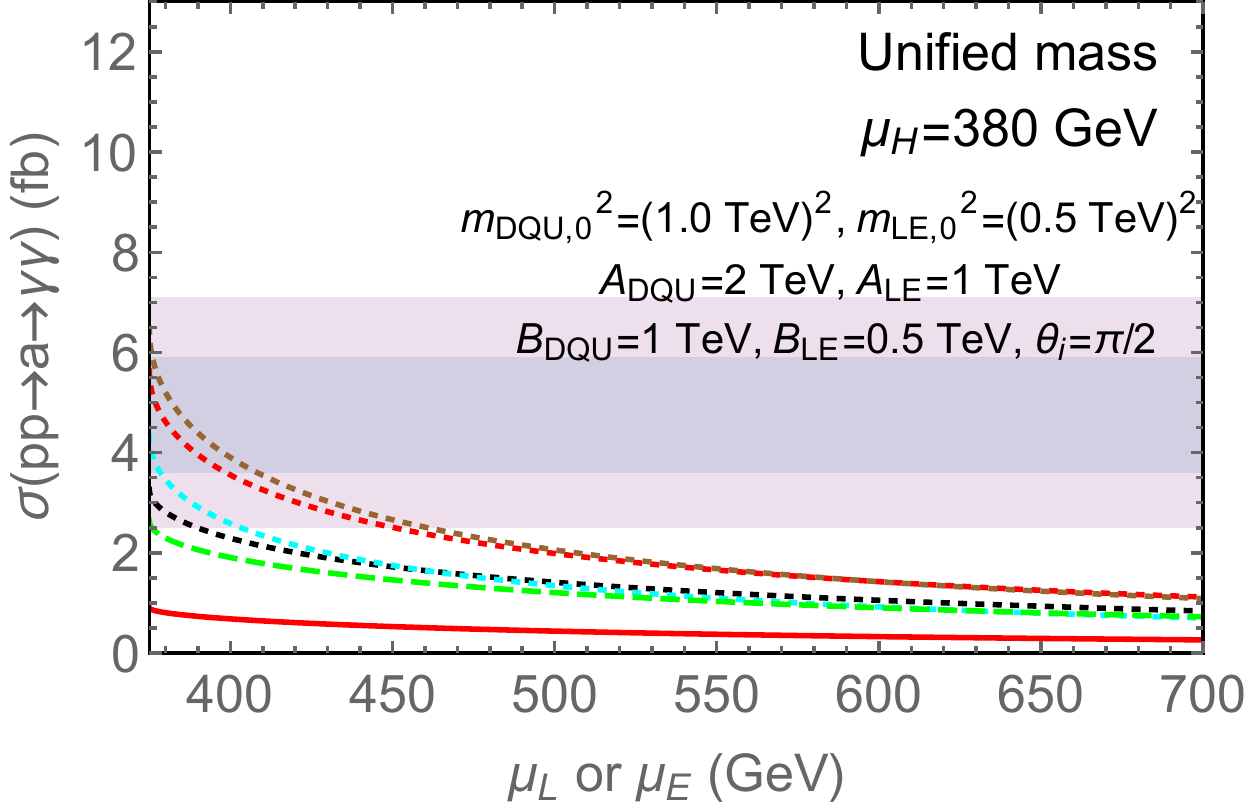}
  \includegraphics[scale=0.6]{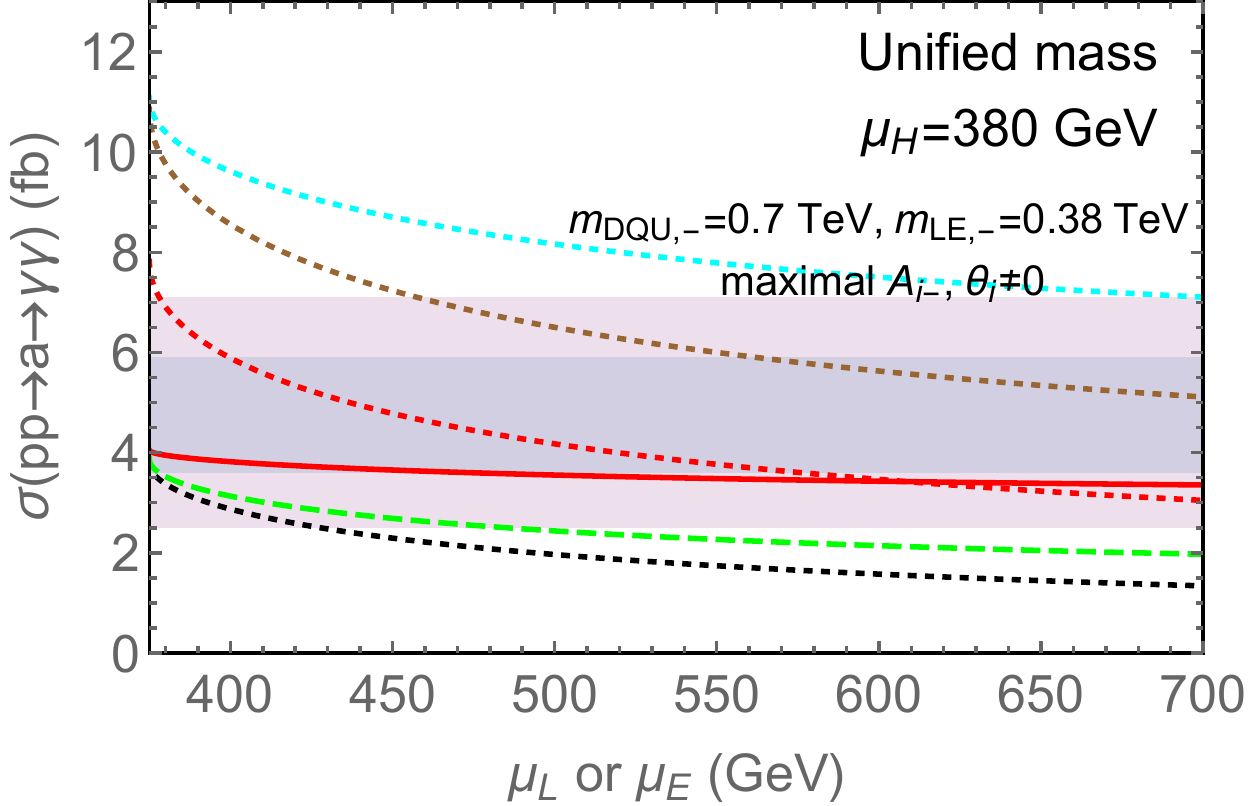}
\end{center}
\caption{Theories with unified mass relations and $\lambda_H \neq 0$: Prediction for $\sigma_a B_{\gamma \gamma}$ at $\sqrt{s} = 13~{\rm TeV}$ as a function of the lightest vector lepton mass, with scalar partners decoupled (upper left panel) and soft masses indicated in the figure (other panels).}
\label{fig:unify_oneres}
\end{figure}

\begin{figure}[t]
\begin{center}
  \includegraphics[scale=0.6]{legend.pdf}\\
  \vspace{0.5cm}
  \includegraphics[scale=0.6]{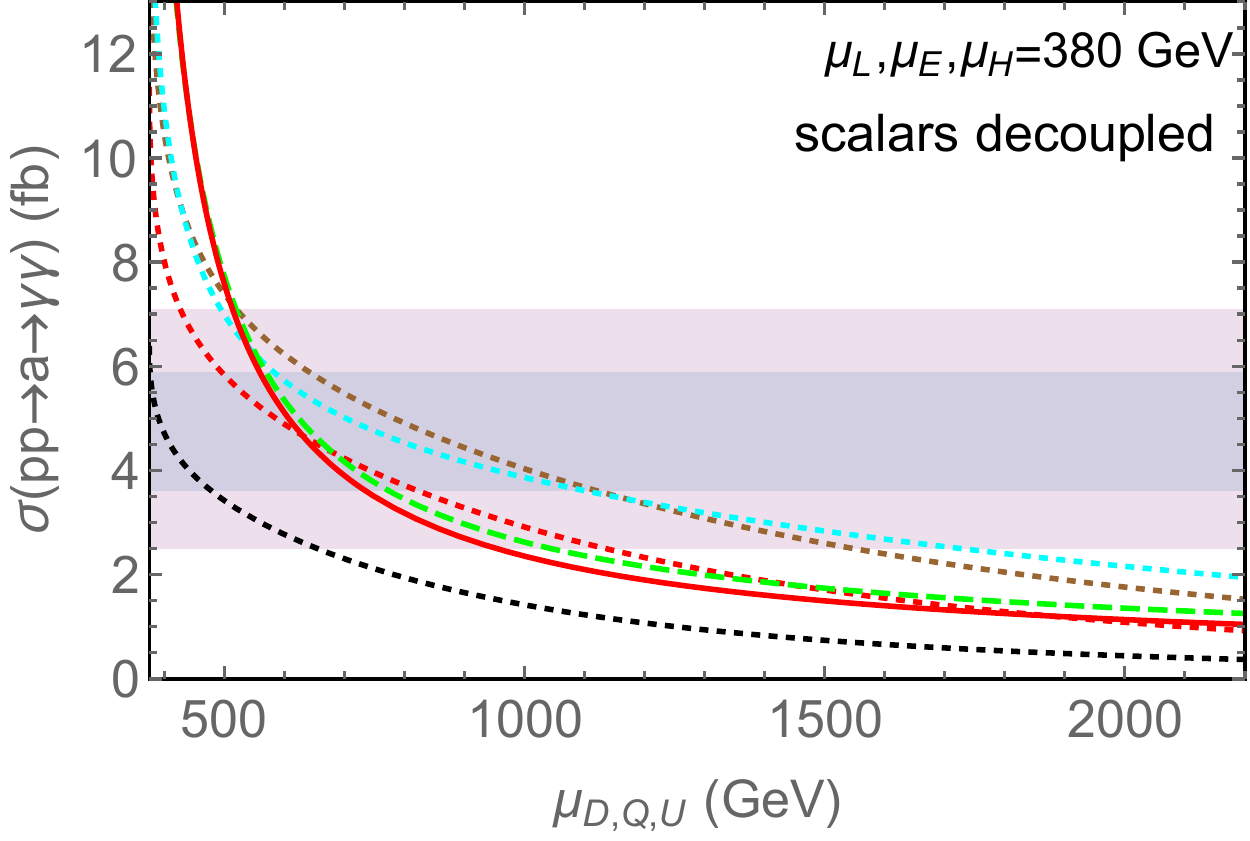}
    \includegraphics[scale=0.6]{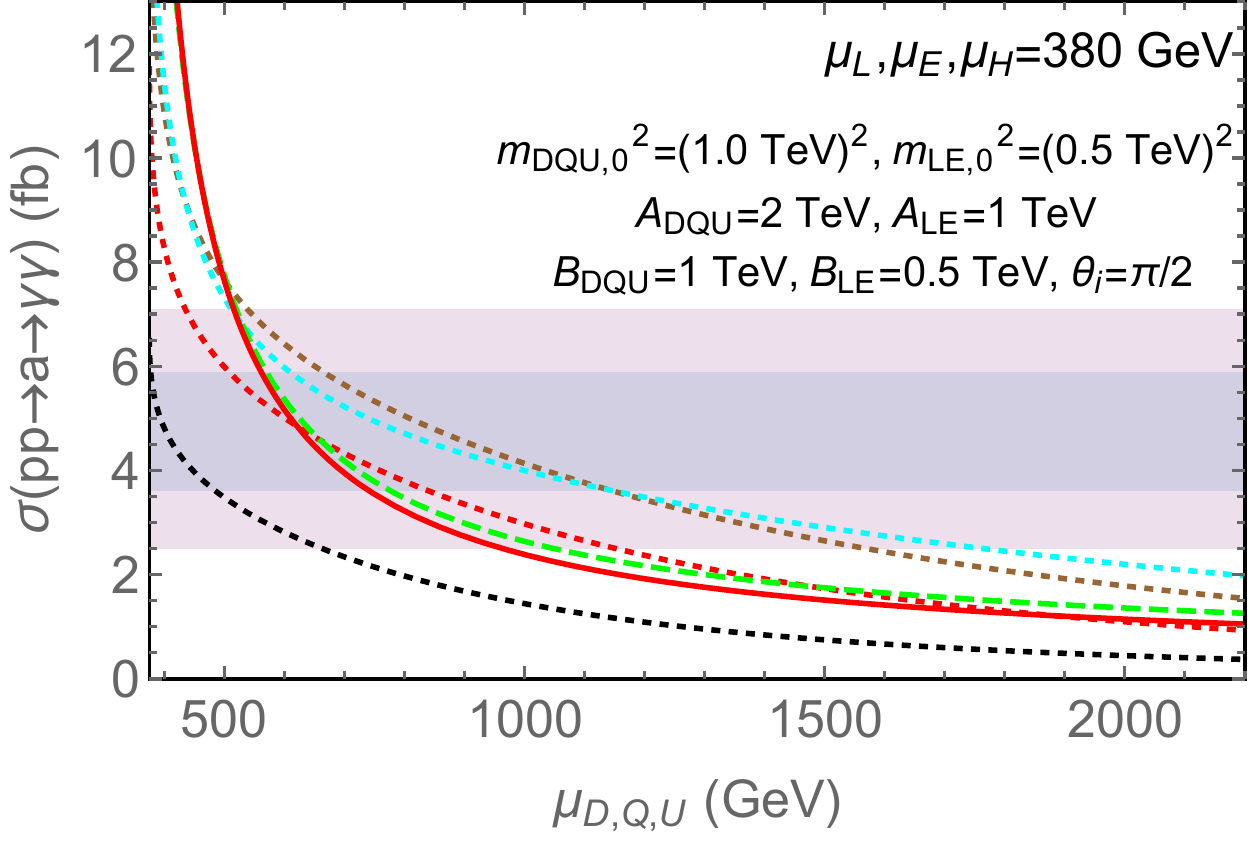}
        \includegraphics[scale=0.6]{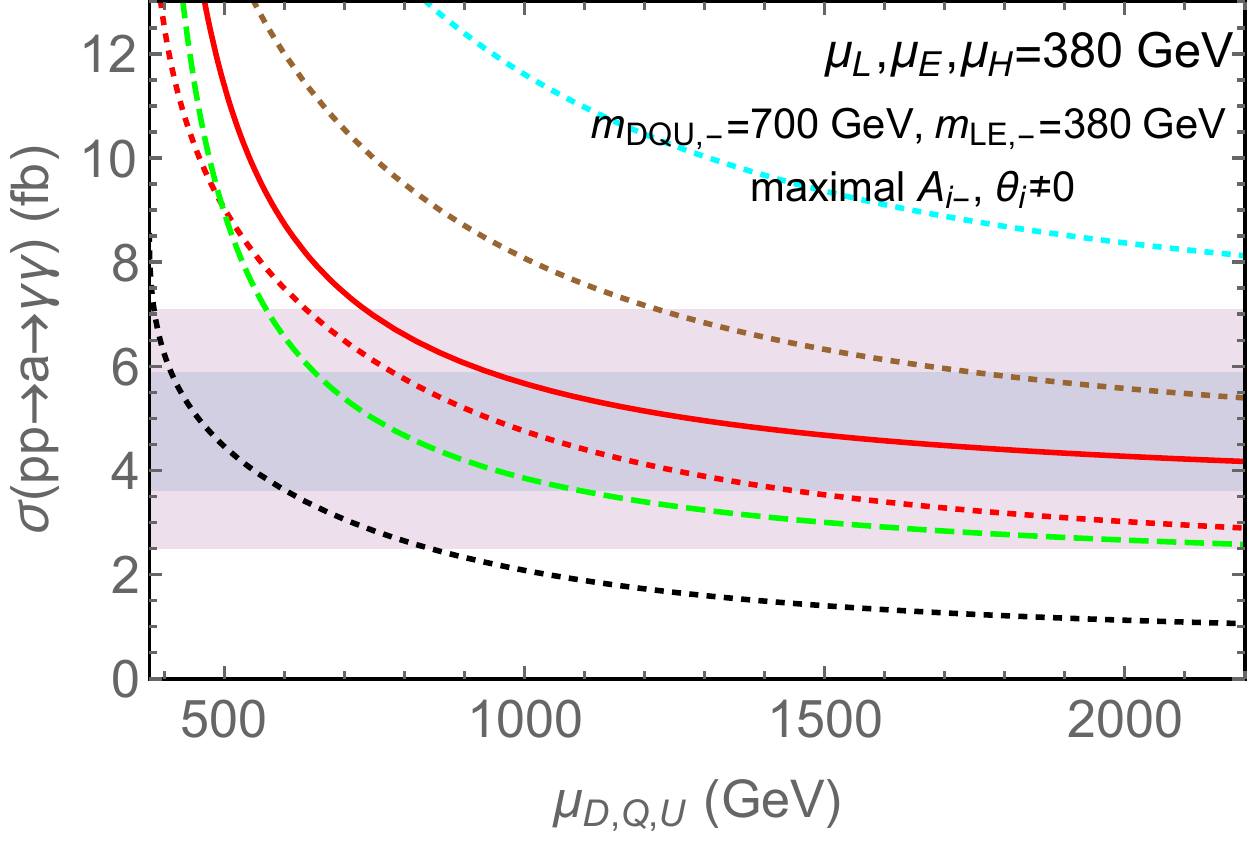}
\end{center}
\caption{Theories without unified mass relations and $\lambda_H \neq 0$: Prediction for $\sigma_a B_{\gamma \gamma}$ at $\sqrt{s} = 13~{\rm TeV}$ as a function of the degenerate vector quark mass, for vector lepton masses and the Higgsino mass at $380~{\rm GeV}$, with scalar partners decoupled (upper left panel) and soft masses indicated in the figure (other panels).}
\label{fig:optimal_oneres}
\end{figure}

The scalar $s$ efficiently decays into the standard model Higgs, $W$ boson, and $Z$ boson, and hence does not contribute to the $750~{\rm GeV}$ excess.  The excess can be still explained by the pseudoscalar $a$.  In Figure~\ref{fig:unify_oneres}, we show the prediction for $\sigma_a {\rm Br}_{\gamma\gamma}$ at the LHC with $\sqrt{s}=13~{\rm TeV}$ as a function of $\mu_L$ for $({\bf 5} + \overline{\bf 5})_i$ and $\mu_E$ for $({\bf 10} + \overline{\bf 10})$ and $({\bf 15} + \overline{\bf 15})$, assuming that $\mu_i$ unify at the unification scale.  In the upper left panel, the contribution from the scalar components of $\Phi_i$ and $\bar{\Phi}_i$ are ignored, while it is taken into account in other panels.  All theories except for $({\bf 15} + \overline{\bf 15})$ can explain the diphoton excess without large $A_i$ terms.  Vector quarks are as heavy as $600-1200~{\rm GeV}$, which is expected to be within the reach of the LHC.  The bound is, however, relaxed by large $A_i$ terms, as shown in the lower panel.  In Figure~\ref{fig:optimal_oneres}, we show the prediction for $\sigma_a {\rm Br}_{\gamma\gamma}$ as a function of degenerate vector quark masses, with vector lepton masses and the Higgsino mass fixed at $380~{\rm GeV}$.  The vector quark masses can be as large as $2~{\rm TeV}$ without large $A_i$ terms.

%

\section{Wide diphoton resonance for {\boldmath $\lambda_H = 0$} and small {\boldmath $B_S \mu_S$}}
\label{sec:width}

In this section, we discuss a possible way to obtain a ``wide width resonance" from the scalar $S$.  As we have pointed out in Ref.~\cite{Hall:2015xds}, the mass difference of a few tens of GeV between the scalar $s$ and the pseudoscalar $a$ can be naturally obtained by a threshold correction at the TeV scale from $\Phi_i$ and $\bar{\Phi}_i$. Then $s$ and $a$ are observed as a single wide resonance.  Here we explore the dependence of the mass splitting on $(A_i, \theta_i)$.

This explanation requires that the holomorphic supersymmetry breaking soft mass of $S$, the $B_S\mu_S$ term, is small.  In gravity mediation, the size of the $B_S$ term is as large as other soft masses, and hence $\mu_S$ should be suppressed.  This requires that the soft mass squared of $S$, $m_S^2$, is positive at the low energy scale. Otherwise, the vev of $S$ is large (see section~\ref{sec:Svev}) and hence fine-tuning is required to obtain small enough $\mu_i$.  In gauge mediation, on the other hand, the $B_S \mu_S$ term is given by a three loop effect and hence is suppressed even if $\mu_S$ is unsuppressed.

\begin{figure}[t]
  \includegraphics[scale=0.6]{legend.pdf}\\
  \vspace{0.5cm}
\begin{tabular}{cc}

\begin{minipage}{0.5\textwidth}
  \includegraphics[scale=0.60]{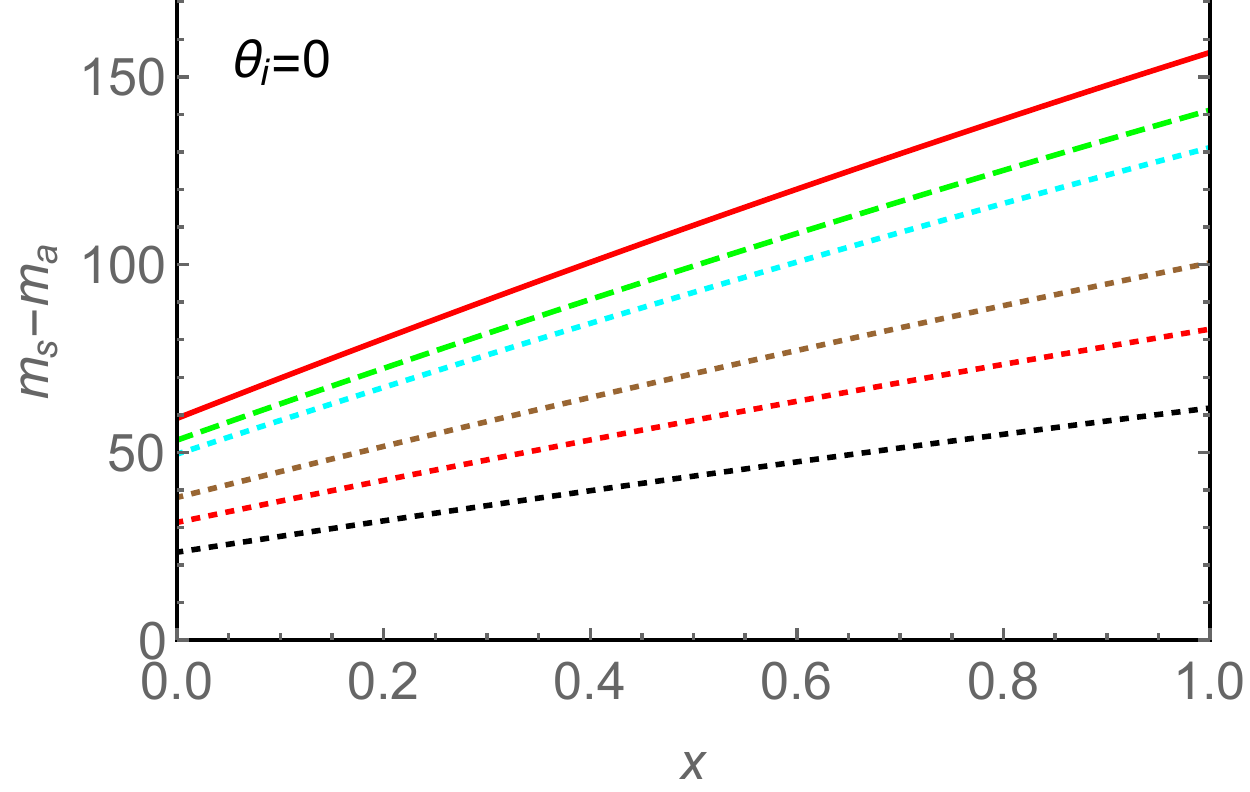}
\end{minipage} 

\begin{minipage}{0.5\textwidth}
\begin{tabular}{|c|c||c|c|}
\hline
$\mu_{DQU}$ & $1~{\rm TeV}$ & $\mu_{LE}$ & $0.4~{\rm TeV}$ \\
$m_{DQU,0}^2$ & $(1~{\rm TeV})^2$ & $m_{LE,0}^2$ & $(0.5~{\rm TeV})^2$ \\
$B_{DQU}$ & $1~{\rm TeV}$ & $B_{LE}$ & $0.5~{\rm TeV}$ \\
$A_{DQU}$ & $2 \times x~{\rm TeV}$ & $A_{LE}$ & $1 \times x~{\rm TeV}$ \\
\hline
\end{tabular}
\end{minipage}\\

\begin{minipage}{0.5\textwidth}
  \includegraphics[scale=0.60]{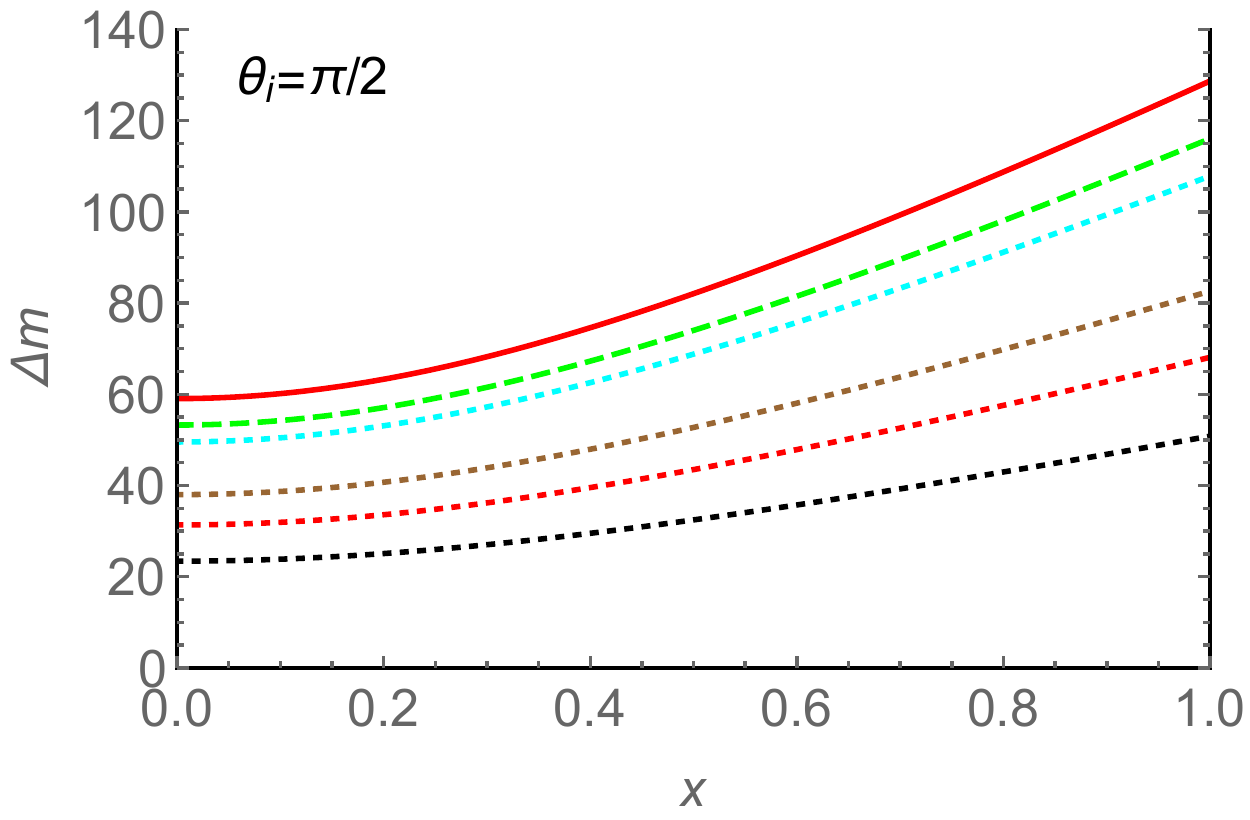}
\end{minipage}

\begin{minipage}{0.5\textwidth}
  \includegraphics[scale=0.60]{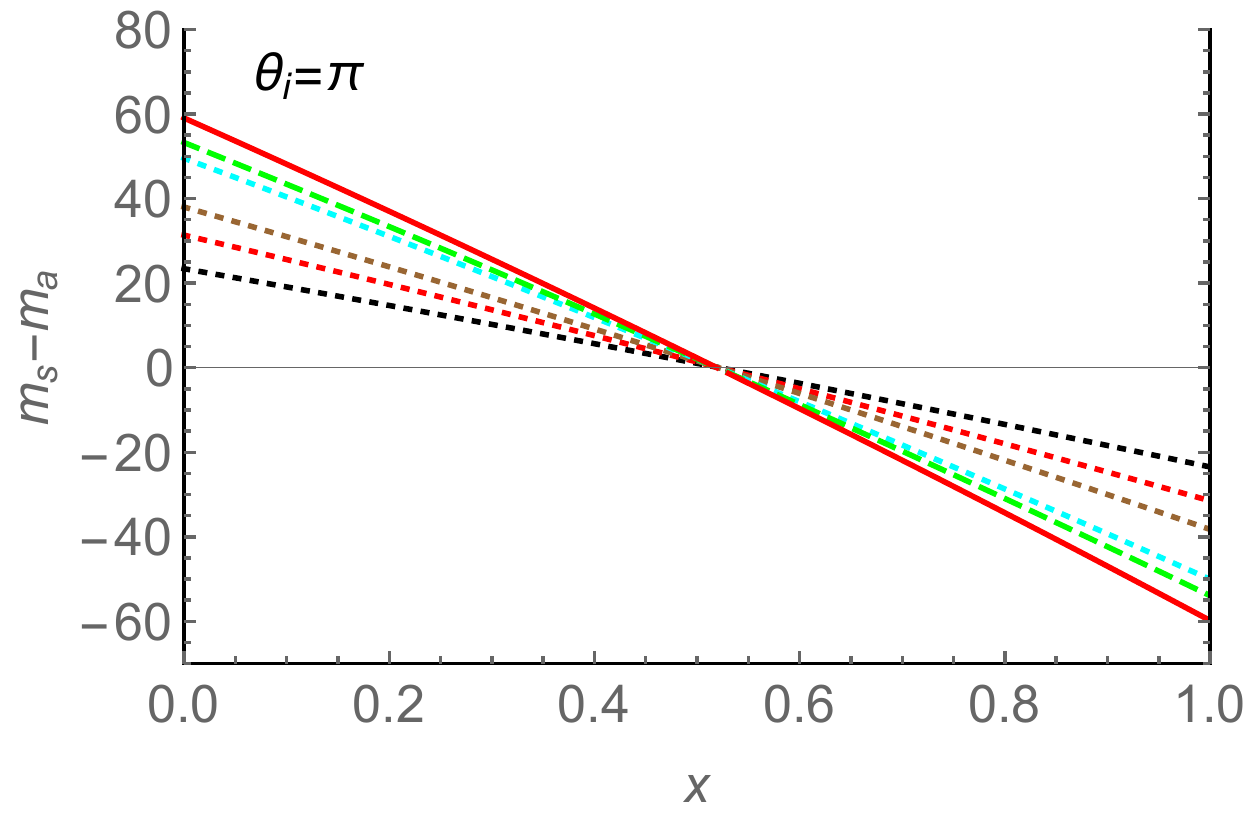}
\end{minipage}

\end{tabular}
\caption{The mass difference between the scalar $s$ and the pseudoscalar $a$ (or the two mass eigenstates of $S$ in the case of CP violation) for three different values of the phase of $B_i \mu_i$.  The horizontal axis, $x$, represents the size of the $A_i$ terms as indicated in the table.}
\label{fig:mass_diff}
\end{figure}

The quantum correction to the mass matrix is given by
\begin{align}
\Delta V =& \frac{1}{2} 
\begin{pmatrix}
s & a
\end{pmatrix}
\begin{pmatrix}
\Delta_{ss} & \Delta_{sa} \\
\Delta_{sa} & 0 
\end{pmatrix}
\begin{pmatrix}
s \\ a
\end{pmatrix}, \\
\Delta_{ss}=& \frac{1}{32\pi^2} \sum_i \lambda_i^2\left[
4 \mu_i^2 {\rm ln} \frac{m_{i+}^2 m_{i-}^2}{\mu_i^4}
+4 \mu_i A_i {\rm cos}\theta_i {\rm ln} \frac{m_{i+}^2}{m_{i-}^2}
+ A_i^2 {\rm cos}2\theta_i \left( 2 - \frac{m_{i+}^2 + m_{i-}^2}{m_{i+}^2 - m_{i-}^2} {\rm ln} \frac{m_{i+}^2}{m_{i-}^2}  \right) \right],\\
\Delta_{sa} =& 
\frac{1}{32\pi^2} \sum_i \lambda_i^2  A_i {\rm sin}\theta_i \left[ 2 \mu_i {\rm ln} \frac{m_{i+}^2}{m_{i-}^2}  + A_i {\rm cos}\theta_i \left( 2 - \frac{m_{i+}^2+m_{i-}^2}{m_{i+}^2-m_{i-}^2} {\rm ln} \frac{m_{i+}^2}{m_{i-}^2} \right)  \right],
\end{align}
where the correction $\Delta_{aa}$ is absorbed into the soft mass squared of $S$.  Note that in the supersymmetric limit, where $A_i=0$ and $m_{i+}=m_{i-}=\mu_i$Ó, the mass difference vanishes.  In Figure~\ref{fig:mass_diff}, the mass difference is shown for each theory as a function of the size of the $A_i$ terms, with the mass parameters shown in the table.  The mass difference can be few tens of GeV.

If $\mu_S=0$, the $\tilde{s}$ mass arises at one loop from virtual vector matter and $\tilde{s}$ may be the lightest supersymmetric particle.  For this to be interpreted as ``singlet-doublet'' dark matter, a mixing with the Higgsino should be introduced.  Further work is needed to investigate whether a small $SH_uH_d$ coupling that provides this mixing also gives a small enough mixing between $s$ and the doublet Higgs boson so that $s$ still contributes to the diphoton resonance.  If so, the predominantly $\tilde{s}$ dark matter may have a mass allowing the observed abundance via freezeout annihilation on the $Z$ or Higgs pole.

\section{Signal of {\boldmath $s$} decay to standard model dibosons}
\label{sec:other scalar}

In this section, we discuss the signal from $s \rightarrow hh, W^+ W^-,ZZ$ at the LHC for $\lambda_H \neq 0$.  The scalar $s$ is produced via gluon fusion and decays into pairs of standard model particles.  If it is heavy enough, it also decays into a pair of vector quarks/leptons.

The mixing between the standard model like Higgs $h$ and the singlet scalar $s$ given by
\begin{align}
\theta_{hs} \simeq \sqrt{2} \lambda_H \frac{v \mu_H}{ m_s^2} = 0.028 \times \frac{\lambda_H}{0.2} \frac{\mu_H}{400~{\rm GeV}} \left(\frac{m_s}{1000~{\rm GeV}}\right)^{-2},
\end{align}
where $m_s$ is the mass of $s$ and $v \simeq 246~{\rm GeV}$ is the vev of the standard model Higgs.  The measurement of the Higgs production cross section restricts the mixing, $\theta_{hs}^2 < 0.1$~\cite{higgscoupling}.  This puts a lower bound on $m_s$,
\begin{align}
m_s > 350~{\rm GeV} \left( \frac{\lambda_H}{0.3}\right)^{1/2} \left( \frac{\mu_H}{380~{\rm GeV}}\right)^{1/2}.
\end{align}

In the limit $m_s \gg m_{h,Z,W}$, the decay width of $s$ into pairs of the standard model Higgs bosons, $W$ bosons, and $Z$ bosons can be evaluated by the equivalence theorem:
\begin{align}
& \Gamma(s\rightarrow hh) \simeq \Gamma(s\rightarrow ZZ) \simeq \frac{1}{2}\Gamma(s\rightarrow W^+W-)
\nonumber \\
& \simeq \frac{\lambda_H^2 }{16\pi} \frac{\mu_H^2}{m_s}  
=0.13~{\rm GeV}\times  \left(\frac{\lambda_H}{0.2}\right)^2 \left(\frac{m_s}{{\rm TeV}}\right)^{-1} \left(\frac{\mu_H}{400~{\rm GeV}}\right)^2.
\end{align}
Through mixing with the standard model Higgs, $s$ decays into a pair of top quarks with a rate
\begin{align}
\Gamma(s\rightarrow t\bar{t}) \simeq \frac{3y_t^2 \lambda_H^2}{16\pi} \frac{v^2 \mu_H^2 }{m_s^3}  
=0.023~{\rm GeV}\times  \left(\frac{\lambda_H}{0.2}\right)^2 \left(\frac{m_s}{{\rm TeV}}\right)^{-3} \left(\frac{\mu_H}{400~{\rm GeV}}\right)^2.
\end{align}
For large $m_s$, the scalar $s$ also decays into a fermionic component of $\Phi_i$ and $\bar{\Phi}_i$ with a decay rate
\begin{align}
\Gamma(s\rightarrow \Phi_i \bar{\Phi}_i) \simeq  \frac{\lambda_H^2 N_i m_s}{16\pi}  \left( 1 - \frac{4 \mu_i^2}{m_S^2}\right)^{3/2}  =1.6~{\rm GeV} \times \left(\frac{\lambda_i}{0.2}\right)^2 \frac{N_i}{2} \frac{m_s}{{\rm TeV}} \left( 1 - \frac{4 \mu_i^2}{m_S^2}\right)^{3/2}.
\end{align}
For simplicity, we assume that the scalar components of $\Phi_i$ and $\bar{\Phi}_i$ are heavy enough that $s$ does not decay into them.  Inclusion of these decay modes is straightforward.

\begin{figure}[t]
\begin{center}
  \includegraphics[scale=0.6]{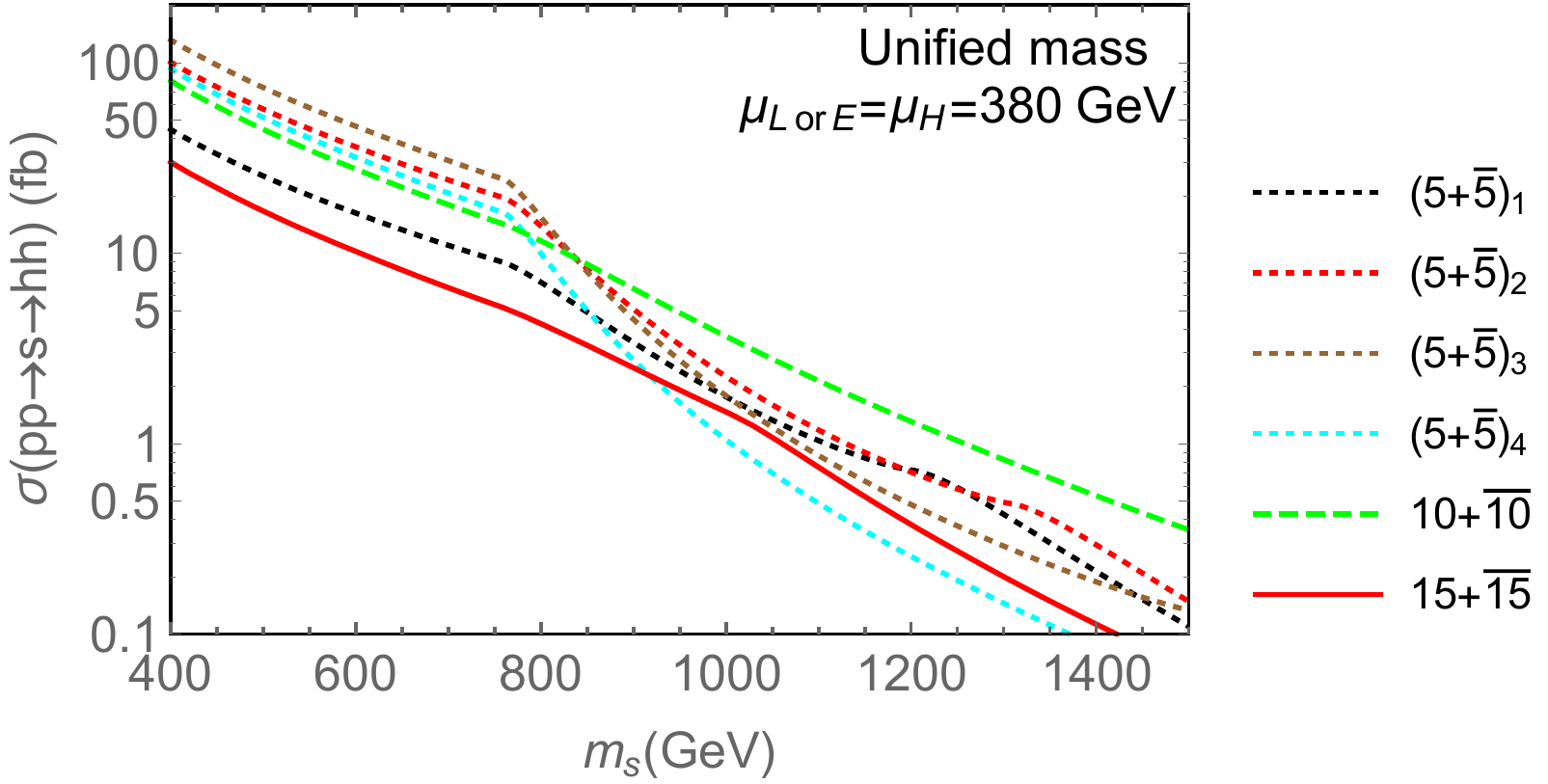}
\end{center}
\caption{Theories with unified mass relations: Prediction for $\sigma_s B_{hh}$ at $\sqrt{s} = 13~{\rm TeV}$ as a function of $m_s$.  Predictions for $\sigma_s B_{WW,ZZ}$ can be estimated by the equivalence theorem.}
\label{fig:unify_s_hh}
\begin{center}
  \includegraphics[scale=0.6]{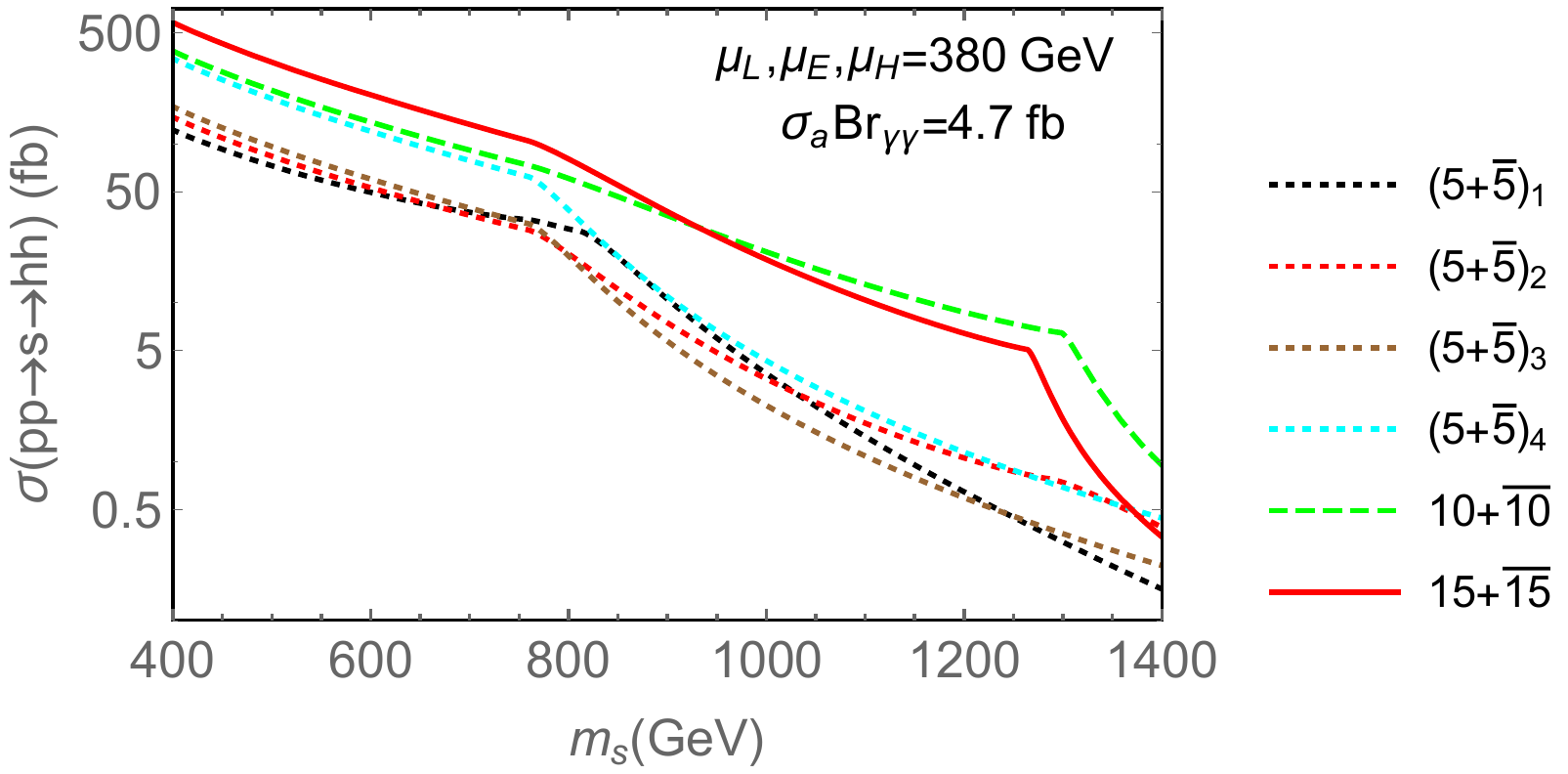}
\end{center}
\caption{Theories without unified mass relations: Prediction for $\sigma_s B_{hh}$ at $\sqrt{s} = 13~{\rm TeV}$ as a function of $m_s$.  Predictions for $\sigma_s B_{WW,ZZ}$ can be estimated by the equivalence theorem.}
\label{fig:optimal_s_hh}
\end{figure}

In Figure~\ref{fig:unify_s_hh}, we show the prediction for $\sigma_s{\rm Br}_{hh}$ at the $13~{\rm TeV}$ LHC as a function of $m_s$,  assuming that $\mu_H$ and the lightest vector-lepton mass ($\mu_L$ for $({\bf 5} + \overline{\bf 5})_i$ and $\mu_E$ for $({\bf 10} + \overline{\bf 10})$ and $({\bf 15} + \overline{\bf 15})$) are $380~{\rm GeV}$ and $\mu_i$ unify at $M_G$.  The signal is depleted for $m_s > 760~{\rm GeV}$ since the decay mode into a pair of vector leptons is open.  In Figure~\ref{fig:optimal_s_hh} we show a similar plot but assuming $\mu_L=\mu_E = \mu_H = 380~{\rm GeV}$ with the masses of the vector quarks determined so that $\sigma_a{\rm Br}_{\gamma\gamma} = 4.7~{\rm fb}$ at the $13~{\rm TeV}$ LHC.  The prediction for $\sigma(pp\rightarrow s \rightarrow WW,ZZ)$ can be estimated by the equivalence theorem.  In both cases, the cross section is predicted to be $O(100-1)~{\rm fb}$ for $m_s = (400-1400)~{\rm GeV}$, which can be tested at the LHC.

\section{Semi-perturbative unification and TeV scale thresholds}
\label{sec:running}

In this section, we discuss gauge coupling unification in $({\bf 5} + \overline{\bf 5})_4$ and $({\bf 15} + \overline{\bf 15})$ theories.  In these theories, gauge couplings $\alpha_i$ become $O(1)$ around the unification scale, and unify in a semi-perturbative regime.  Nevertheless, as we will show, precision gauge coupling unification is successfully achieved with moderate threshold corrections around the TeV scale.

\begin{figure}[t]
\begin{center}
    \includegraphics[scale=0.6]{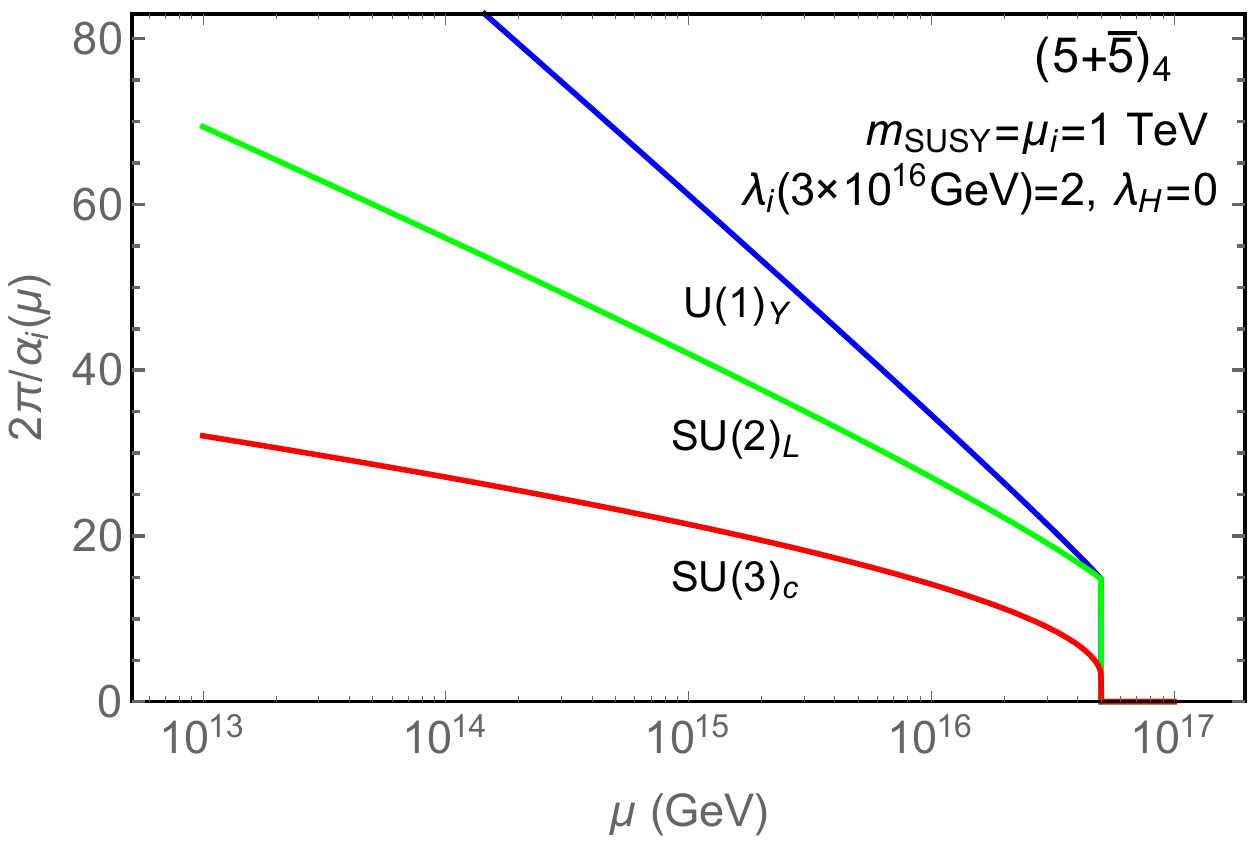}
  \includegraphics[scale=0.6]{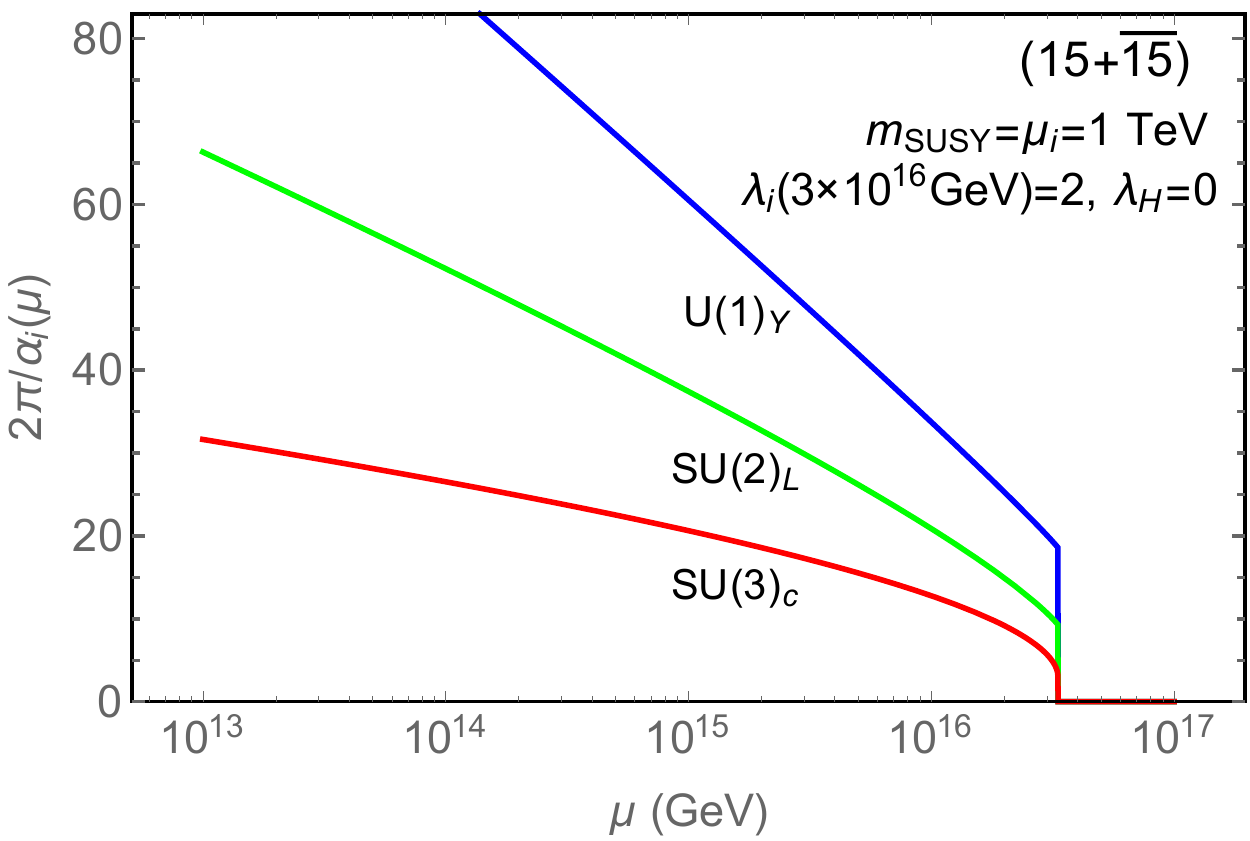}
\end{center}
\caption{Running of the standard model gauge couplings for $({\bf 5} + \overline{\bf 5})_4$ and $({\bf 15} + \overline{\bf 15})$, with masses of all MSSM particles and vector quarks/leptons of $1~{\rm TeV}$.}
\label{fig:gauge_run_up}
\end{figure}

In Figure~\ref{fig:gauge_run_up}, we show the running of the standard model gauge couplings for $({\bf 5} + \overline{\bf 5})_4$ and $({\bf 15} + \overline{\bf 15})$ with the NSVZ beta function~\cite{Novikov:1983uc}, evaluating anomalous dimensions at the one-loop level.  Here we assume that the masses of all MSSM particles and vector quarks/leptons are $1~{\rm TeV}$.  It can be seen that the $SU(3)_c$ gauge coupling enters the non-perturbative regime before unification.  The perturbative unification of gauge couplings requires large threshold corrections at a high energy scale or smaller threshold corrections at the TeV scale.

\begin{figure}[t]
\begin{center}
 \includegraphics[scale=0.7]{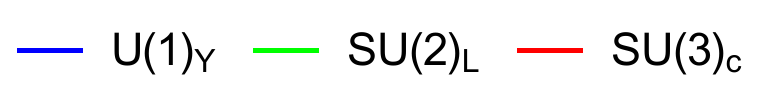} \\
 \includegraphics[scale=0.6]{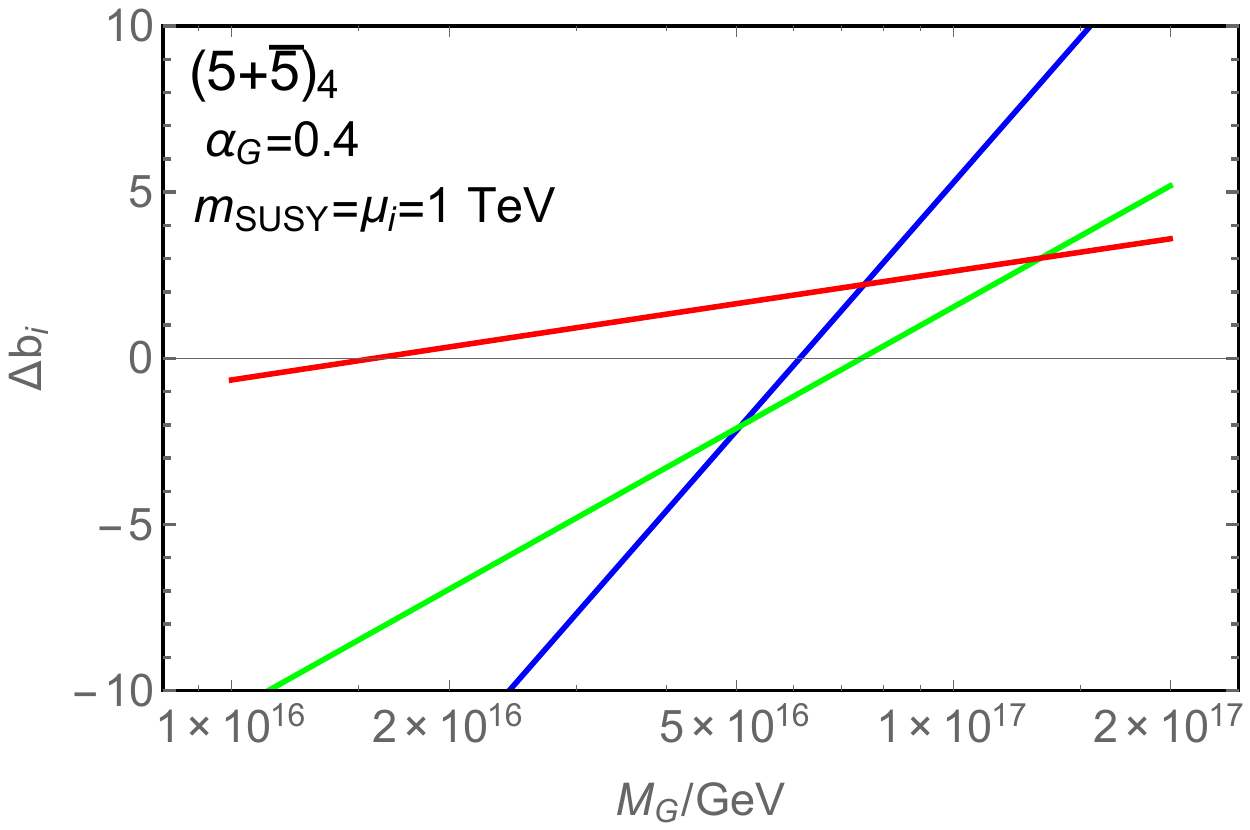}
 \includegraphics[scale=0.6]{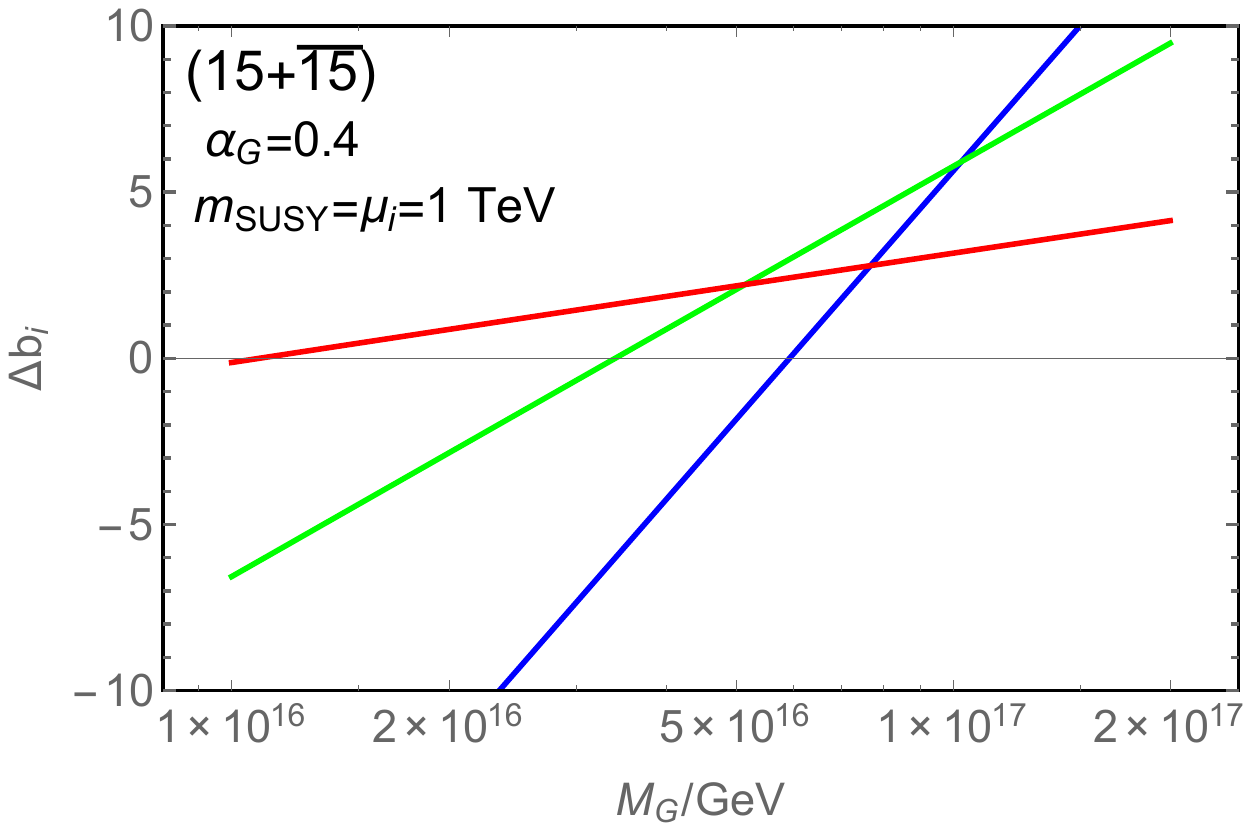}
 \includegraphics[scale=0.6]{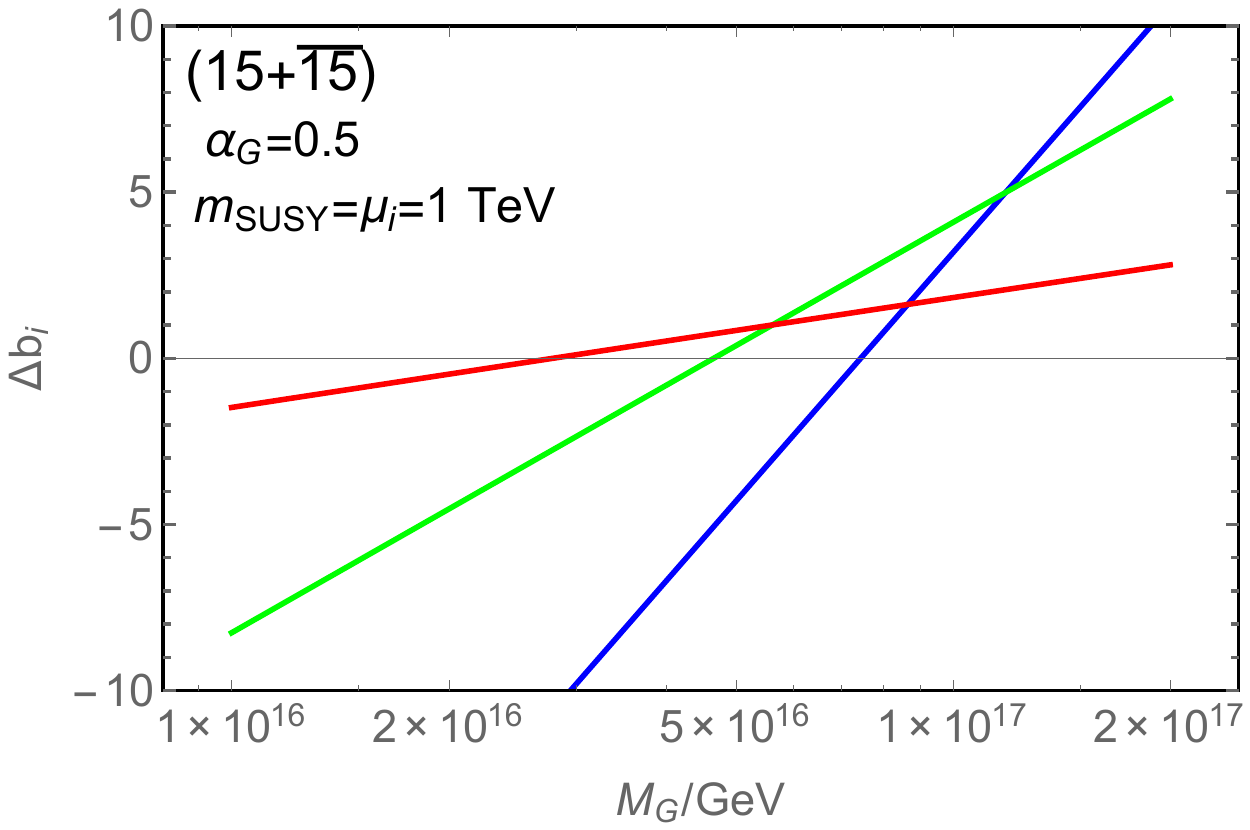}
 \includegraphics[scale=0.6]{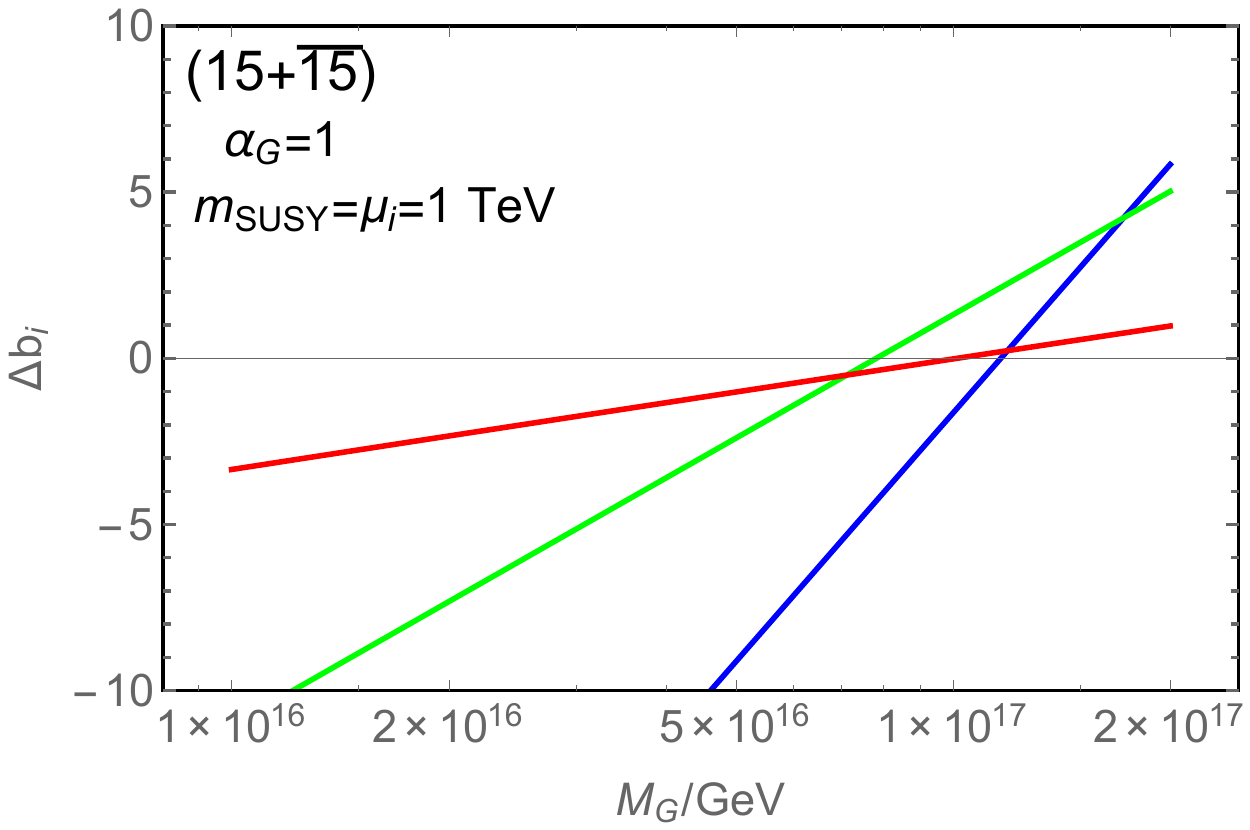}
\end{center}
\caption{The difference between the predicted and observed gauge couplings at the weak scale as a function of the unification scale $M_G$, with various $\alpha_G$.  Here, the masses of all MSSM particles and vector quarks/leptons are $1~{\rm TeV}$ and $\lambda(M_G)=2$.}
\label{fig:gauge_run_down}
\end{figure}

To assess the required threshold corrections at the TeV scale, we solve the renormalization group equation from the unification scale down to the electroweak scale.  In Figure~\ref{fig:gauge_run_down}, we show $\varDelta b_i$, the difference between the predicted and observed gauge couplings at the weak scale
\begin{align}
\varDelta b_i \equiv \frac{2\pi}{\alpha_i(m_Z)}\biggr|_{\rm prediction} - \frac{2\pi}{\alpha_i(m_Z)}\biggr|_{\rm observed},
\end{align}
as a function of the unification scale $M_G$, with various $\alpha_G$.  Here we assume that the masses of all MSSM particles and vector quarks/leptons are $1~{\rm TeV}$ and $\lambda_i(M_G)=2$.

In each panel of Figure~\ref{fig:gauge_run_down}, the couplings come close to unifying in the region of $M_G \sim (5 \times 10^{16}~\mbox{--}~10^{17})~{\rm GeV}$, where $\varDelta b_i$ are all positive and typically $3~\mbox{--}~5$.  These are not very large and hence can be countered by TeV scale threshold corrections.  As superpartner and/or vector quark and lepton masses are increased above $1~{\rm TeV}$, the predicted gauge couplings at $M_Z$ become larger and hence the lines in Figure~\ref{fig:gauge_run_down} are lowered, so that raising these masses produces threshold corrections of the required sign.  For precision unification the three curves must intersect at a point where $\varDelta b_i=0$.  For $({\bf 15} + \overline{\bf 15})$, this means that the curve for $SU(2)$ must be lowered more than the curve for $SU(3)$---the masses for particles with $SU(2)_L$ charge must be raised further than the masses for colored particles.

\begin{figure}[t]
\begin{center}
  \includegraphics[scale=0.7]{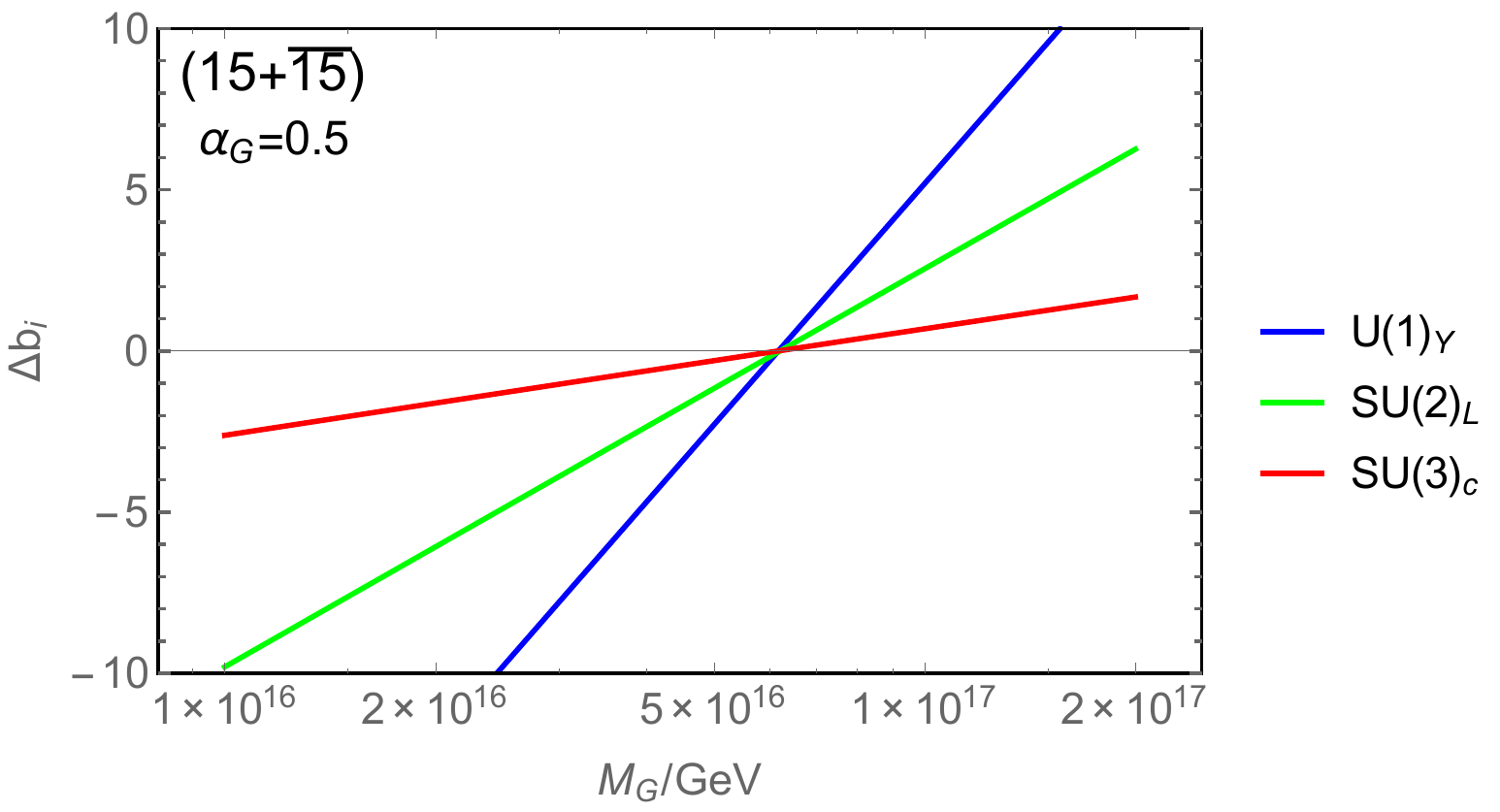}
\end{center}
\caption{Precision gauge coupling unification in $({\bf 15} + \overline{\bf 15})$ with masses of superparticles and vector quarks/leptons shown in Table~\ref{tab:sample}.}
\label{fig:gauge_run_sample}
\end{figure}

Such a mass spectrum is difficult to achieve in conventional supersymmetric unification scenarios, where boundary conditions at the unification scale and renormalization running typically lead to colored particles heavier than non-colored particles.  Precision unification in $({\bf 15} + \overline{\bf 15})$ calls for a non-conventional scenario, such as unified symmetry breaking by boundary conditions in extra dimensions.  For example, the masses of superparticles and vector quarks/leptons in Table~\ref{tab:sample}, with wino heavier than gluino, predict gauge couplings at $M_Z$ in agreement with the observed values, as shown in Figure~\ref{fig:gauge_run_sample}.
\begin{table}[htdp]
\caption{A sample mass spectrum of MSSM superparticles (upper row) and vector quarks/leptons (lower row).  Here, $m_{QUDLE} \equiv \sqrt{m_{QUDLE+} m_{QUDLE-}}$.  With this mass spectrum, the prediction for the gauge couplings at the weak scale is improved, as is shown in Figure~\ref{fig:gauge_run_sample}.}
\begin{center}
\begin{tabular}{|c|c|c|c|c|c|c|c|c|c|}
\hline
$m_{\tilde{q}}$ & $m_{\tilde{u}}$ & $m_{\tilde{d}}$ & $m_{\tilde{l}}$ & $m_{\tilde{e}}$ & $m_H$ & $\mu_H$ & $m_{\tilde{g}}$ & $m_{\tilde{w}}$ & \\ \hline
1600 & 1200 & 1200 & 1500 & 600 & 2000 & 380 & 1500 & 4000 &  \\ \hline \hline
$m_{Q}$ & $m_{U}$ & $m_{D}$ & $m_{L}$ & $m_{E}$ & $\mu_Q$ & $\mu_U$ & $\mu_D$ & $\mu_L$ & $\mu_E$ \\ \hline
1500 & 1000 & 1000 & 1000 & 500 & 800 & 800 & 800 & 380 & 380 \\ \hline
\end{tabular}
\end{center}
\label{tab:sample}
\end{table}
%

\section{Vacuum expectation value for {\boldmath $S$} and soft operators}
\label{sec:Svev}

\subsection{Vacuum expectation value for {\boldmath $S$}}

In estimating the diphoton signal rate, we assumed that the phases of $\lambda_i$ are aligned with each other in the basis where $\mu_i$ have a common phase.  Even with CP conservation in the superpotential, we have assumed that the signs of $\lambda_i \mu_i$ are independent of $i$.  This alignment maximizes the diphoton rate and, while it is not necessary for large $A_i$ and non-unified masses, in other cases it is helpful in obtaining a sufficient diphoton rate.  This alignment is naturally achieved if $\mu_i$ are forbidden by some symmetry, under which $S$ is charged, and are solely given by the vev of $S$.  Thus, instead of Eq.~(\ref{eq:superpotential}) we may start with the much simpler superpotential
\begin{align}
W \supset  S \sum_i \lambda_i \Phi_i \overline{\Phi}_i  + \lambda_H S H_u H_d.
\label{eq:superpotential2}
\end{align}
In this case $\mu_i = \lambda_i \vev{S}$ and the spectrum of vector matter is given by the ``Unified" case, with the diphoton rate given in Figure~\ref{fig:unify_twores}.  In the absence of large $A_i$, the upper panels show that only the $({\bf 5} + \overline{\bf 5})_{2,3,4}$ theories explain the diphoton resonance.  However, the lower panels show that the scalar contribution with large $A_i$ allows all theories to explain the diphoton resonance.  After electroweak symmetry breaking the soft trilinear scalar interaction proportional to $A_H$ leads to a linear term in $S$, which will therefore develop a vev.  However, even in the large $A_H$ limit this is too small to give sufficient mass to the vector matter.  Some other origin for a large vev must be found.

One idea for achieving this is to give a negative mass squared to $S$, with a restoring term in the potential for $S$ arising from the superpotential coupling $\kappa$ of Eq.~(\ref{eq:superpotential}).  Assuming that $B_S\mu_S$ is negligible, as occurs if $\mu_S \propto \vev{S}$, the vev of $S$ is given by
\begin{align}
|\vev{S}| = \frac{1}{\kappa} \sqrt{m_s^2 + \frac{m_a^2}{3}} > 2200~{\rm GeV} \frac{0.2}{\kappa},
\label{eq:svev}
\end{align}
where we have used $m_s = 0$ and $m_a = 750~{\rm GeV}$ to obtain the last inequality.  The coupling $\kappa$, however, receives large renormalization and its size at the low energy scale is much smaller than the one at the unification scale.  In $({\bf 5} + \overline{\bf 5})$ theory, $\kappa({\rm TeV})=0.3$ for $\lambda_i(M_G)=1$ and $\kappa(M_G)=3$.  The corresponding lower bound on $\mu_L$(TeV) is $940~{\rm GeV}$, which is too large to explain the diphoton excess.  In other theories, the lower bound is severer.  Theories with $\mu_{i,H,S}$ generated from $\vev{S}$ can explain the diphoton signal if the superpotential couplings $\lambda_{i,H,S}$ become strong at scales of $(10 - 10^3)~{\rm TeV}$, since then $\kappa$(TeV) can be sufficiently large~\cite{Barbieri:2016cnt}.  However, for perturbative couplings to the unification scale, $\mu_i$ cannot arise from $\vev{S}$ of Eq.~(\ref{eq:svev}).

Another possibility is that $\vev{S}$ arises from a positive mass squared and a tadpole term.  One may wonder whether the mechanism to yield the tadpole term in general generates $\mu_i$ terms independent of the vev of $S$.  This is avoided by the so-called SUSY-zero.  Consider, for example, an $R$ symmetry with a charge assignments $S(-2)$ and $\Phi_i\bar{\Phi}_i(4)$.  (Construction of a similar mechanism with a non-$R$ symmetry is straightforward.)  In any supersymmetric theory, the superpotential, which has an $R$ charge of $2$, must have a non-zero vev to cancel the cosmological constant induced by supersymmetry breaking.  We denote the chiral operator of $R$ charge $2$ that condenses and generates the superpotential vev as ${\cal O}$.  The tadpole term of $S$ is given by
\begin{align}
K = {\cal O} S + {\rm h.c.}.
\end{align}
In gravity mediation, this term generates a tadpole term $\sim({\rm TeV})^3 S$ and hence $\vev{S}=O(1)~{\rm TeV}$.  On the other hand, the superpotential term $W\sim {\cal O}\Phi_i \bar{\Phi}_i$ is forbidden (except for $Z_{4R}$).  It is essential that there is no chiral operator, $\bar{{\cal O}}$, having $R$ charge $-2$ and a similar expectation value as ${\cal O}$; otherwise, the superpotential term $W\sim \bar{\cal O} \Phi_i \bar{\Phi}_i$ generates $\mu_i$ independent of $\vev{S}$.  Such a chiral operator is actually absent when $R$ symmetry is broken by a gaugino condensation.  This mechanism leads to Eq.~(\ref{eq:superpotential2}) with $S$ having a vev of order the supersymmetry breaking scale, providing the messenger scale of order the Planck mass.

The $R$ symmetry forbids both $S^2$ and $S^3$ interactions, so that at tree-level $\mu_S=0$.  (For a discrete $Z_6$ $R$ symmetry $S^2$ is allowed. In this case, the degeneracy of $s$ and $a$ cannot be naturally explained.)  The fermionic component of $S$, $\tilde{s}$, is massless at tree-level and is expected to be the lightest supersymmetric particle.  Since the $R$ symmetry is broken by the supersymmetry breaking interactions, the $\tilde{s}$ mass appears at the TeV scale from integrating out $\Phi_i$ and $\bar{\Phi}_i$ at one loop.  These radiative contributions to $m_{\tilde{s}}$ are proportional to $\mu_i$ and $B_i\mu_i$, and are of order $O(10-100)~{\rm GeV}$ for soft masses of a TeV scale, suggesting that predominantly $\tilde{s}$ neutralino dark matter results from annihilation via the $Z$ or Higgs pole.

\subsection{The scale of soft operators and fine-tuning}

Consider the mass scale of the soft supersymmetry breaking at low energies.  For a fixed value of the gaugino masses, for example close to the experimental limit, as more vector quarks/leptons are added to the theory, the gaugino mass at the unification scale becomes larger for a high messenger scale.  This raises the overall soft mass scale for the scalar superpartners,  leading to fine-tuning to obtain a singlet scalar at $750~{\rm GeV}$ and scalar vector quarks/leptons sufficiently light to contribute to the diphoton signal.  For $({\bf 5} + \overline{\bf 5})_{4}$, $({\bf 10} + \overline{\bf 10})$ and $({\bf 15} + \overline{\bf 15})$ theories, the required fine-tuning to obtain the $750~{\rm GeV}$ mass amounts to $O(1)$\%.  The fine-tuning is severer, typically by a factor of 10, if the mass squared of $S$, $m_S^2$, at the TeV scale is required to be positive (see sections~\ref{sec:width} and \ref{sec:Svev}).  This is because the renormalization of soft masses makes $m_S^2$ negative at the TeV scale, unless $m_S^2$ is positive and large at the unification scale.  To avoid the tachyonic masses of the vector squarks/sfermions, their soft masses must be also large enough at a high energy scale, which raises soft mass scales further.  Such fine-tuning can be avoided by introducing non-standard low scale mediation of supersymmetry breaking or non-standard running of soft operators, for example induced by conformal hidden sector interactions~\cite{Luty:2001jh,Dine:2004dv,Murayama:2007ge}.  Such a conformal sector also has the potential to yield large $A$ terms~\cite{Murayama:2007ge}, which are favored by the diphoton signal and the Higgs mass of $125~{\rm GeV}$.

In section~\ref{sec:running} we found that, for precision gauge coupling unification in $({\bf 15} + \overline{\bf 15})$, non-standard soft operators at the TeV scale were also required.

\section{Summary and discussion}

Following an initial study in Ref.~\cite{Hall:2015xds}, we confirm that the reported $750~{\rm GeV}$ diphoton resonance can be explained by supersymmetric theories that add a gauge singlet $S= (s + ia) / \sqrt{2}$ and vector matter $(\Phi_i, \bar{\Phi}_i)$ to the minimal set of particles:  there are 6 possibilities for vector matter that allow perturbative gauge coupling unification, and the case of a full generation of vector matter is particularly interesting as it leads to $\alpha_G \sim 1$.  For each of these 6 possibilities there are two versions of the theory with different Higgs phenomenology, depending on whether $\lambda_H S H_u H_d$ is included.  For $\lambda_H =0$ $(\neq 0)$ the theory should be viewed as vector matter added to the MSSM (NMSSM).  

For $\lambda_H \neq 0$, a narrow $750~{\rm GeV}$ resonance arises from $a \rightarrow \gamma \gamma$ and we predict a second resonance decaying to dibosons $s \rightarrow hh, ZZ, W^+ W^-$, with a rate typically accessible in future LHC runs as shown in Figures~\ref{fig:unify_s_hh} and \ref{fig:optimal_s_hh}.  For $\lambda_H = 0$, there is no mixing of $s$ with the Higgs boson so there are two diphoton resonances arising from $a,s \rightarrow \gamma \gamma$.  If one of these produces the observed resonance at $750~{\rm GeV}$, the other may be of much higher mass, and both would be narrow.  Alternatively, if the mass splitting between $s$ and $a$ is small they may both contribute to the observed diphoton signal, leading to an apparent width of order the mass splitting.

The diphoton event rate depends on several factors: the quantum numbers and masses of the vector quarks and leptons, the masses of the vector squarks and sleptons (which depend on $A$ parameters and phases), whether the vector quark and lepton masses obey unified  relations, and whether the resonance is produced by $a$, $s$ or both.  For unified vector quark and lepton mass relations and decoupled vector squarks and sleptons the event rate is sufficient only for $({\bf 5} + \overline{\bf 5})_{2,3,4}$ theories, whether $\lambda_H$ is zero or not; and even these theories require vector lepton masses below $(400-450)~{\rm GeV}$.  The rate is substantially increased by having non-unified vector quark and lepton masses and by including contributions from vector squark and slepton loops, as shown in Figures~\ref{fig:unify_twores}~--~\ref{fig:optimal_oneres}.  By comparing the upper and lower panels of these figures one sees that the largest increase in the diphoton signal results from allowing large $A$ terms~\cite{Nilles:2016bjl}.  Indeed, maximal values of $A$ consistent with vacuum stability allow vector quarks to be decoupled in some theories, with the signal arising from vector leptons, sleptons and squarks.  However, these $A$ terms are very large and, for moderate values of $A$ in the $1-2~{\rm TeV}$ range, the vector quarks are predicted to lie within the LHC reach, as shown by the upper right panels of Figures~\ref{fig:unify_twores}~--~\ref{fig:optimal_oneres}.

The diphoton event rate also depends on whether the amplitudes from the various vector matter multiplets add coherently.  This occurs automatically if the vector matter masses arise from $S$ acquiring a vev.  In section~\ref{sec:Svev} we introduce a theory with an $R$ symmetry that accomplishes this in a way that explains why all the superpotential mass parameters have a scale governed by supersymmetry breaking.

There is an interesting possibility that for $\lambda_H = 0$ the mass splitting between $s$ and $a$ arises dominantly from loops of vector matter~\cite{Hall:2015xds}.  In Figure~\ref{fig:mass_diff} we extend our analysis to show that the corresponding width of the diphoton resonance is sensitive to $A$ terms and CP violating phases. 

While perturbative supersymmetric unified theories can easily account for the diphoton signal, we find it likely that some scheme beyond gravity mediation is needed for soft operators and their running.  The extra matter makes the gluino mass very large at unification scales which then typically leads to masses for the scalar superpartner that are too large.  This problem is strengthened as more vector multiplets are added, and in the $({\bf 5} + \overline{\bf 5})_4$ and $({\bf 15} + \overline{\bf 15})$ theories we also find that for the gauge couplings to remain perturbative we need either non-standard boundary conditions or running of the soft parameters.  Furthermore, in the theory introduced to align the phases of the amplitudes for the diphoton resonance, vacuum stability also suggests non-standard running of soft operators.

\section*{Acknowledgments}
This work was supported in part by the Department of Energy, Office of Science, Office of High Energy Physics, under contract No. DE-AC02-05CH11231, by the National Science Foundation under grants PHY-1316783 and PHY-1521446, and by MEXT KAKENHI Grant Number 15H05895.  LJH thanks the Simons Foundation and the Institute for Theoretical Studies, ETH, Zurich.

\appendix

\section{Maximal {\boldmath $A$} Terms}
\label{sec:A term}

In this appendix, we estimate the bound on the size of the $A_i$ terms from vacuum stability.  We consider cases with $\theta_i \simeq 0,\pi$. (The constraint for other $\theta_i$ can be obtained by taking into account the appropriate factors of $\cos\theta_i$ and $\sin\theta_i$ as well as the dynamics of $a$).  The strongest constraint comes from the tunneling involving $\Phi_{i-}$ and $s$.  Hereafter we drop the subscripts $i$ and $-$.  The scalar potential of $\Phi$ and $s$ is given by
\begin{align}
V(s, \Phi) = \frac{\lambda^2}{4}|\Phi|^4 + \frac{\lambda^2}{2} s^2 |\Phi|^2 - \frac{\lambda}{\sqrt{2}} A s |\Phi|^2 + \frac{1}{2}m_s^2 s^2 + m_\Phi^2 |Q|^2.
\end{align}
We take $A > 0$ without loss of generality.  After a change of variables, $s \rightarrow m_s \sigma / \lambda$, $\Phi \rightarrow m_s \phi / \sqrt{2}\lambda$ and $x^\mu \rightarrow \xi^\mu / m_s$, the action is given by
\begin{equation}
\lambda^2 S = \int d^4 \xi \left[ \frac{1}{2} \partial \sigma \partial \sigma + \frac{1}{2} \partial \phi \partial \phi  - {\cal V} (\sigma,\phi)  \right],
\end{equation}
where
\begin{equation}
{\cal V} = \frac{1}{16} \phi^4 + \frac{1}{4} \phi^2 \sigma^2 - \frac{r_A}{2 \sqrt{2}} \phi^2 \sigma + \frac{1}{2}\sigma^2 + \frac{1}{2} r_\phi^2 \phi^2,~~r_A \equiv \frac{A}{m_s},~~r_\phi \equiv \frac{m_\phi}{m_s}.
\end{equation}

We consider the tunneling path with the minimum potential barrier, in which
\begin{align}
\sigma = \frac{r_A}{2\sqrt{2}}\frac{\phi^2}{1 + \phi^2/2}.
\end{align}
Along this path, the potential is given by
\begin{align}
{\cal V}_{\rm eff}(\phi) = \frac{1}{16} \phi^4 + \frac{1}{2} r_\phi^2 \phi^2 - \frac{r_A^2}{16} \frac{\phi^4}{ 1 + \phi^2/2},
\label{eq:potential}
\end{align}
and the canonically normalized field is given by
\begin{align}
\phi_c \equiv \sqrt{1 +  \frac{r_A^2}{2} \frac{\phi^2}{ (1+ \phi^2/2)^2}} \phi.
\end{align}
\begin{figure}[t]
\begin{center}
  \includegraphics[scale=0.7]{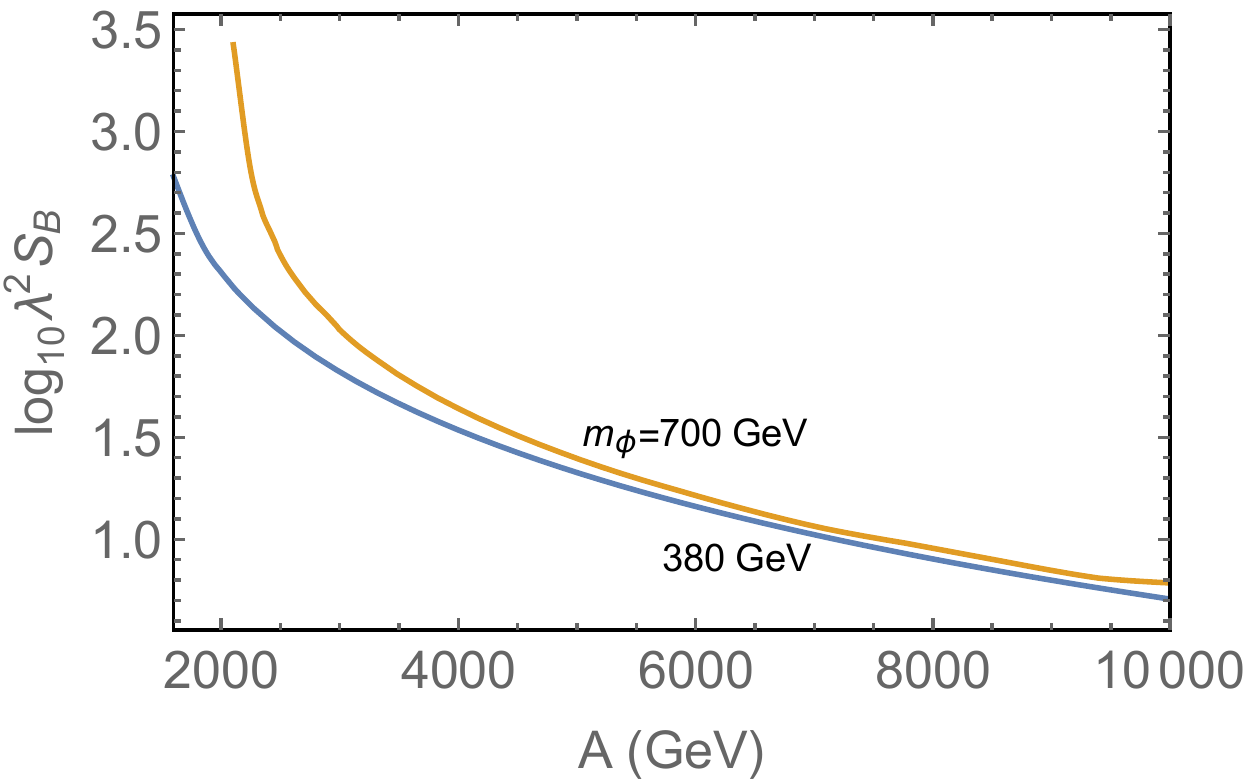}
\end{center}
\caption{The bounce action as a function of $A$ for $m_\phi = 380~{\rm GeV}$ and $700~{\rm GeV}$.}
\label{fig:bounce action}
\end{figure}
We numerically obtain the bounce action~\cite{Coleman:1977py} solving the equation of motion of $\phi_c$.  In Figure~\ref{fig:bounce action}, we show the size of the bounce action, $S_B$, as a function of $A$ for $m_\phi = 380~{\rm GeV}$ and $700~{\rm GeV}$.  We require that the lifetime of the vacuum is longer than the age of the universe, $S_B>400$.  In Table~\ref{tab:A term}, we show the upper bound on $A_{i-}$ for each theory.  The result is consistent with the one presented in Ref.~\cite{Nilles:2016bjl} within a few tens of percent.
\begin{table}[htdp]
\caption{The upper bound on $A_{i-}$ in GeV, taking $m_{DQU,-}=700~{\rm GeV}$ and $m_{LE,-}=380~{\rm GeV}$.}
\begin{center}
\begin{tabular}{|c|c|c|c|c|c|c| c|}
\hline
$\lambda_H=0$ &$({\bf 5} + \overline{\bf 5})$&$({\bf 5} + \overline{\bf 5})_2$&$({\bf 5} + \overline{\bf 5})_3$&$({\bf 5} + \overline{\bf 5})_4$& $({\bf 10} + \overline{\bf 10})$ & $({\bf 15} + \overline{\bf 15})$ \\ \hline
$A_{D-}$ & 3100  &3700  & 4000 & 4200 && 4700  \\
$A_{L-}$ & 2200  & 2700 & 3300 & 5200 & & 6700  \\
$A_{Q-}$  & && & & 3300 & 3400 \\
$A_{U-}$  & & && & 4000 & 4400 \\
$A_{E-}$  &&&&&4500& 9400 \\ \hline
\end{tabular}
\begin{tabular}{|c|c|c|c|c|c|c| c|}
\hline
$\lambda_H \neq 0$ &$({\bf 5} + \overline{\bf 5})$&$({\bf 5} + \overline{\bf 5})_2$&$({\bf 5} + \overline{\bf 5})_3$&$({\bf 5} + \overline{\bf 5})_4$& $({\bf 10} + \overline{\bf 10})$ & $({\bf 15} + \overline{\bf 15})$ \\ \hline
$A_{D-}$ & 3100  &3800  & 4200 & 4200 && 4700  \\
$A_{L-}$ & 2300  & 2900 & 3600 & 5500 & & 7100  \\
$A_{Q-}$  & && & & 3400 & 3400 \\
$A_{U-}$  & & && & 4100 & 4400 \\
$A_{E-}$  &&&&&4500& 9400 \\ \hline
\end{tabular}
\end{center}
\label{tab:A term}
\end{table}

\end{document}